\documentclass[aps,prb,showpacs,amsmath,amssymb,reprint,10pt,superscriptaddress,longbibliography]{revtex4-2}
\usepackage[colorlinks=true,urlcolor=blue,citecolor=blue]{hyperref}
\usepackage{url}

\usepackage[english]{babel}
\usepackage[utf8]{inputenc}
\usepackage{graphicx,epsfig,xcolor}
\usepackage{latexsym}
\usepackage{amsfonts}
\usepackage{amsmath}
\usepackage{placeins}
\onecolumngrid
\usepackage{polski}
\usepackage{braket}
\usepackage[bottom]{footmisc}
\usepackage[normalem]{ulem}

\newcommand{\Tr}{\mathrm{Tr}}
\def\ket|#1>{| #1 \rangle}
\def\bra<#1|{\langle #1 |}
\def\<{\langle}
\def\>{\rangle}
\def\{{\lbrace}  
\def\}{\rbrace}
\def\beq{\begin{equation}}
\def\eeq{\end{equation}}

\begin{document}

\title{Quantifying non-Hermiticity using single- and many-particle quantum properties}
\author{Soumik Bandyopadhyay}
\email{soumik.bandyopadhyay@unitn.it}
\affiliation{Pitaevskii BEC Center, CNR-INO and Dipartimento di Fisica, Universit\`a di Trento, Via Sommarive 14, Trento, I-38123, Italy}
\affiliation{INFN-TIFPA, Trento Institute for Fundamental Physics and Applications, Trento, Italy}
\author{Philipp Hauke}
\email{philipp.hauke@unitn.it}
\affiliation{Pitaevskii BEC Center, CNR-INO and Dipartimento di Fisica, Universit\`a di Trento, Via Sommarive 14, Trento, I-38123, Italy}
\affiliation{INFN-TIFPA, Trento Institute for Fundamental Physics and Applications, Trento, Italy}
\author{Sudipto Singha Roy}
\email{sudipto@iitism.ac.in}
\affiliation{Pitaevskii BEC Center, CNR-INO and Dipartimento di Fisica, Universit\`a di Trento, Via Sommarive 14, Trento, I-38123, Italy}
\affiliation{INFN-TIFPA, Trento Institute for Fundamental Physics and Applications, Trento, Italy}
\affiliation{Department of Physics, Indian Institute of Technology (ISM) Dhanbad, IN-826004, Dhanbad, India}

\date{\today}
\begin{abstract}
The non-Hermitian paradigm of quantum systems displays salient features drastically different from Hermitian counterparts. In this work, we focus on one such aspect, the difference of evolving quantum ensembles under $H_{\mathrm{nh}}$ (right ensemble) versus its Hermitian conjugate, $H_{\mathrm{nh}}^{\dagger}$ (left ensemble). We propose a formalism that quantifies the (dis-)similarity of these right and left ensembles, for single- as well as many-particle quantum properties. Such a comparison gives us a scope to measure the extent to which non-Hermiticity gets translated from the Hamiltonian into physically observable properties. We test the formalism in two cases: First, we construct a non-Hermitian Hamiltonian using a set of imperfect Bell states, showing that the non-Hermiticity of the Hamiltonian does not automatically comply with the non-Hermiticity at the level of observables. 
Second, we study the interacting Hatano--Nelson model with asymmetric hopping as a paradigmatic quantum many-body Hamiltonian. Interestingly, we identify situations where the measures of non-Hermiticity computed for the Hamiltonian, for single-, and for many-particle quantum properties behave  distinctly from each other. Thus, different notions of non-Hermiticity can become useful in different physical scenarios.
Furthermore, we demonstrate that the measures can qualitatively mark the model's Parity--Time (PT) symmetry-breaking transition. Our findings can be instrumental in unveiling new exotic quantum phases of non-Hermitian quantum many-body systems as well as in preparing resourceful states for quantum technologies.   
\end{abstract}
\maketitle

\begin{figure}
  \includegraphics[width=\linewidth]{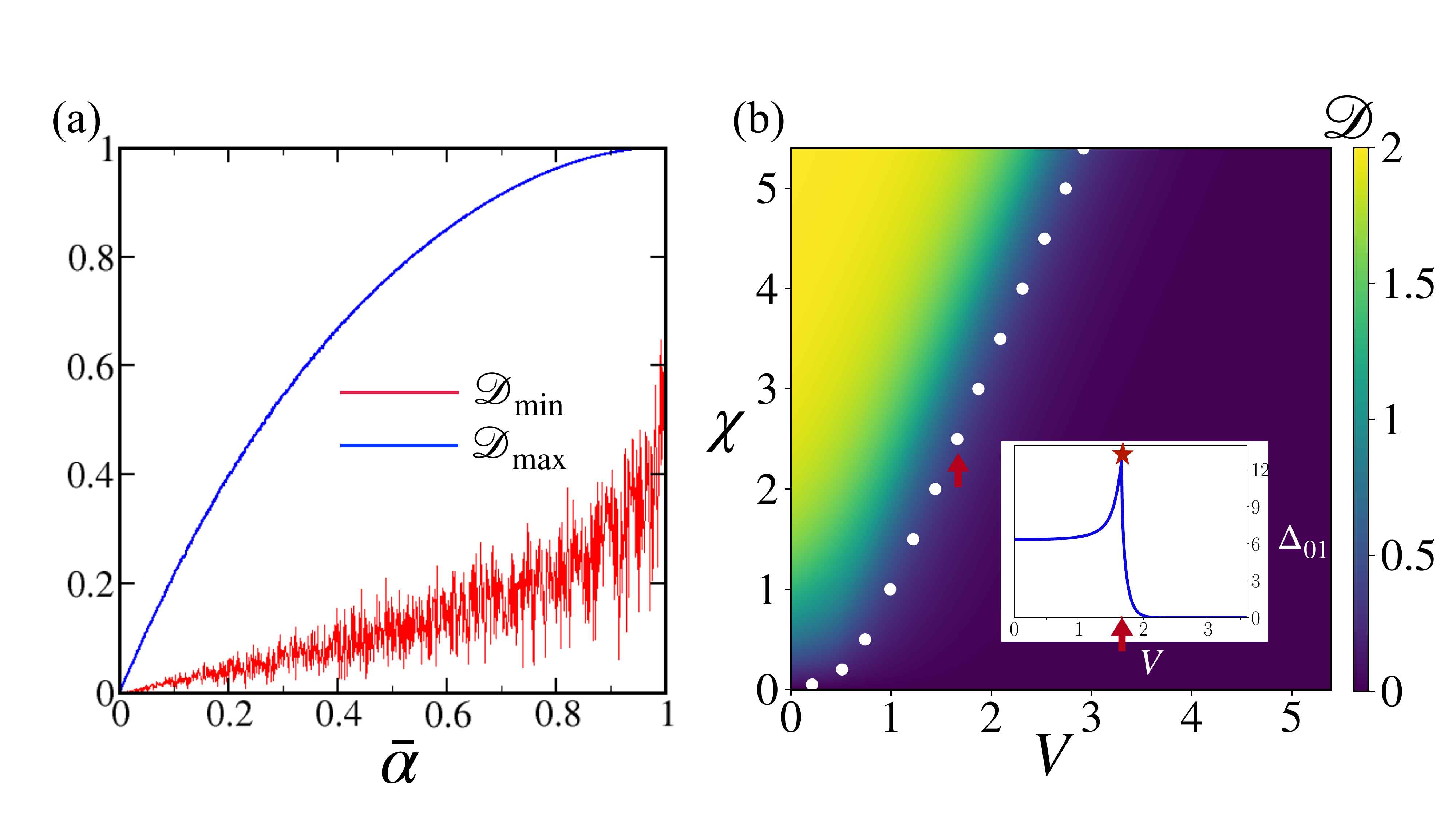}
 \caption{Behavior of degree of non-Hermiticity computed using $\mathcal{D}=\frac{||\hat{H}_{\mathrm{nh}}-\hat{H}_{\mathrm{nh}}^\dagger||}{||\hat{H}
 _{\mathrm{nh}}||}$ for two different non-Hermitian Hamiltonians. (a) Behavior of $\mathcal{D}$ for $\hat{H}^{\mathrm{Bell}}_{\mathrm{nh}}$ constructed using a set of imperfect Bell states as defined in Eqs.~(\ref{eqn:right_vec}) and (\ref{eqn:left_vec}). 
 For each value of the non-Hermiticity parameter $\bar{\alpha}=1-\alpha$, we plot the maximum  ($\mathcal{D}_{\max}$, blue) and the minimum  ($\mathcal{D}_{\min}$, red) within 1000 random realizations of the Hamiltonian obtained by choosing its energy eigenvalues $\{\lambda_m \}$ from a Gaussian distribution with zero mean and unit variance.   
 While  $\mathcal{D}_{\min}$ increases only slowly and under strong fluctuations, $\mathcal{D}_{\max}$ exhibits a clear monotonic growth  with $\bar{\alpha}$. 
 (b) Behavior of $\mathcal{D}$ for the interacting non-Hermitian Hatano--Nelson model  ($\hat{H}^{\mathrm{HN}}_{\mathrm{nh}}$).  
 At low values of interaction $V$, $\mathcal{D}$ increases monotonically with asymmetric coupling $\chi$ and saturates to $\mathcal{D}\approx 2$. 
 However, even for moderately large values of $V$, $\mathcal{D}$ takes a significant non-zero value only beyond a threshold coupling, $\chi_c$. Further, beyond  $V\gtrsim 10^4$,  $\mathcal{D}$ remains zero for all values of  $\chi$ considered in our analysis. 
 In other words, under strong interactions the distance between $\hat{H}_{\mathrm{nh}}^{\mathrm{HN}}$ and $(\hat{H}_{\mathrm{nh}}^{\mathrm{HN}})^\dagger$ becomes insignificant in comparison to the norm of $\hat{H}_{\mathrm{nh}}^{\mathrm{HN}}$ (see also Fig.~\ref{fig:Ham_unnorm} in Appendix~\ref{appendixC}). The white circles correspond to the PT symmetry-breaking transition of the model marked by non-analytic behavior of the finite-size level-spacing $\Delta_{01}:=\mathrm{Re}(\lambda_1 - \lambda_0)$ (inset: $\Delta_{01}$ versus $V$ for fixed $\chi=2.5$). There is a qualitative agreement between the regions where $\mathcal{D}$ drops and non-analyticities in $\Delta_{01}$. The data is reported for $N = 12$ sites, half-filling, and with anti-periodic boundary conditions. Note that in (b), and in all subsequent plots
where the parameter $V$ is considered, it is presented on a logarithmic
scale.
 }
  \label{fig:Ham}    
\end{figure}
\section{introduction}
    The non-Hermitian paradigm of quantum systems, though often challenging our common understanding and interpretation of properties related to physical systems, may appear naturally as a result of valid physical processes. For instance, it may emerge via ubiquitous effects of noise in real quantum systems \cite{photonic1,photonic2,ueda} or constrained dynamics realized via suppressing the quantum jumps in dissipative quantum dynamics \cite{nh_quantumjump1,nh_quantumjump2,nh_review,Nori}. 
    In recent years,  many concepts of Hermitian systems have been generalized to the non-Hermitian context, including  bulk-boundary correspondence \cite{nh_bulk_boundary},  topological characterization of non-Hermitian systems \cite{topology_nh1,topology_nh2}, eigenstate thermalization \cite{ETH_nh_ours,ETH_nh}, many-body localization \cite{Nh_MBL}, etc. Moreover, apart from fundamental aspects, recent works also highlight promising applications of the non-Hermitian framework in the context of quantum technologies, such as in  quantum thermal machines exploiting exceptional points \cite{Nh_thermodyn} or enhanced quantum sensing \cite{Nh_sensing1,Nh_sensing2,Nh_sensing3,Nh_sensing4,Nh_sensing5}. 
Despite this importance, many fundamental aspects of non-Hermitian systems are still unclear. For instance, for a non-Hermitian system there exist different choices of eigenbasis (right, left, or mixed basis), which can yield results that lack a proper physical interpretation \cite{Bardarson,Ryu, Modak}. 
Even more, there is no clear framework how to quantify, using physically observable properties, the non-Hermiticity of a given system.

Here, we present such a framework. 
To provide a concise overview, we briefly anticipate its main features, while we discuss its details in the bulk of the article. 
As is well known, the left and right eigenvectors diagonalizing a non-Hermitian Hamiltonian $\hat{H}_{\mathrm{nh}}=\sum_m \lambda_m |R_m\rangle \langle L_m|$,  obeying $\hat{H}_{\mathrm{nh}}|R_m\rangle=\lambda_m |R_m\rangle$, $\langle L_m|\hat{H}_{\mathrm{nh}}=\langle L_m| \lambda_m$, and $\langle L_m| R_n \rangle =\delta_{mn}$,  can in general be distinct, $\langle L_m|\neq (|R_m\rangle)^\dagger$. 
As we discuss, the evolution generated by $\hat{H}_{\mathrm{nh}}$ leads an initial state to ensembles of right eigenvectors (including, e.g., single eigenstates reachable at large times), $\rho_{RR}(t)=\frac{e^{-it\hat{H}_{\mathrm{nh}}}\rho_{\mathrm{in}} e^{it\hat{H}^\dagger_{\mathrm{nh}}}}{\mathrm{Tr}[e^{-it\hat{H}_{\mathrm{nh}}}\rho_{\mathrm{in}} e^{it\hat{H}^\dagger_{\mathrm{nh}}}]}$, such that observables measured on the system will physically be of the form $\langle R|\hat{O}|R \rangle$. 
In contrast, the evolution governed by $\hat{H}_{\mathrm{nh}}^\dagger$ leads to ensembles of left eigenvectors, $\rho_{LL}(t)=\frac{e^{-it\hat{H}^\dagger_{\mathrm{nh}}} \rho_{\mathrm{in}} e^{it \hat{H}_{\mathrm{nh}}}}{\mathrm{Tr}[e^{-it\hat{H}^\dagger_{\mathrm{nh}}} \rho_{\mathrm{in}} e^{it\hat{H}_{\mathrm{nh}}}]}$, and thus to expectation values of the form $\langle L|\hat{O}|L \rangle$. 
As we further show, the dynamics under {\it any} non-Hermitian Hamiltonian  $\hat{H}_{\mathrm{nh}}$ can be modeled as an effect of complete quantum measurements and subsequent post-selection of the results, giving a clear physical meaning to these expectation values in terms of observable properties. 
Unlike the Hermitian case, the evolution governed by  $\hat{H}_{\mathrm{nh}}$ and $\hat{H}_{\mathrm{nh}}^\dagger$  may lead to results that can significantly differ from each other, due to a rich interplay between potentially complex $\lambda_m$ and/or the difference between $\{|R_m \rangle\}$ and $\{\langle L_m|\}$. 
Our aim is to design measures that systematically capture these differences between $\hat{H}_{\mathrm{nh}}$ and $\hat{H}_{\mathrm{nh}}^\dagger$, and thus quantify the non-Hermiticity of the system through observable properties. 

As a first measure, we introduce the ``Hamiltonian non-Hermiticity'' $\mathcal{D}=\frac{||\hat{H}_{\mathrm{nh}}-\hat{H}_{\mathrm{nh}}^\dagger||}{||\hat{H}_{\mathrm{nh}}||}$, where $||.||$ is the operator norm. This quantifier is a fundamental measure for non-Hermiticity, but it relies on knowledge of the full model Hamiltonian. Even when it is possible to engineer $H$ and $H^\dagger$ and both are theoretically known, one may perhaps not fully trust the precision of the engineering or it may not be straightforward to numerically compute $\mathcal{D}$. 
In order to make non-Hermiticity more easily accessible, we {\color{red} thus} further define the ``non-Hermiticity score" at the level of observables as ${SC}^\mathcal{F}_{\mathrm{nh}}=|\mathcal{F}_{RR}[\rho_{RR}]-\mathcal{F}_{LL}[\rho_{LL}]|$. This score quantifies the non-Hermiticity at the level of any physical state-dependent function $\mathcal{F}[.]$ related to the model by measuring the difference when evaluating $\mathcal{F}[.]$ for the state evolved under $\hat{H}_{\mathrm{nh}}$ versus the one evolved under $\hat{H}_{\mathrm{nh}}^\dagger$. 

Besides presenting complementary possibilities to certify non-Hermiticity, the 
above formalism opens up the possibility to analyze to what extent the degree of non-Hermiticity of the Hamiltonian $\hat{H}_{\mathrm{nh}}$ gets translated to the behavior of physically observable quantities.
Moreover, we find situations where the Hamiltonian non-Hermiticity is small (due to a term with a large norm) while the non-Hermiticity score of simple observables is large, and vice versa. Thus, different non-Hermiticity quantifiers complement each other, being able to reveal different notions of non-Hermiticity. Furthermore, as we demonstrate in our work, the scores may become instrumental in marking non-Hermitian phase transitions.

The rest of this article is organized as follows. In Sec.~\ref{sec:theory}, 
introduce the quantifiers ``Hamiltonian non-Hermiticity" as well as the ``non-Hermiticity score."  In Sec.~\ref{sec:models}, we illustrate the framework, first at the example of a non-Hermitian Hamiltonian constructed using a set of imperfect Bell states. Subsequently, we discuss the behavior in the interacting Hatano--Nelson model with asymmetric hopping, for which we compare non-Hermiticity of the model Hamiltonian and the non-Hermiticity score for single- as well as multi-site properties. Finally, in Sec.~\ref{sec:conclusions}, we conclude with a brief discussion.

\section{Theoretical background}
\label{sec:theory}

In this section, we  introduce two (nonequivalent) ways of quantifying non-Hermiticity through the full Hamiltonian or through observables. 

\subsubsection{Hamiltonian non-Hermiticity}
As a first measure, we introduce the ``Hamiltonian non-Hermiticity" (see also Fig.~\ref{fig:Ham}).  
It puts the qualitative notion of comparing a system Hamiltonian and its Hermitian conjugate onto quantitative grounds.  
We define it as 
\begin{equation}
    \label{eq:Hnhscore}
    \mathcal{D}=\frac{||\hat{H}_{\mathrm{nh}}-\hat{H}_{\mathrm{nh}}^\dagger||}{||\hat{H}_{\mathrm{nh}}||}\,.
\end{equation}
In this paper, we use the operator norm (which for a given operator $\hat{X}$ is defined as the square root of the largest eigenvalue  ($\lambda_{\text{max}}$) of $ \hat{X}^\dagger \hat{X}$, i.e., 
$
\|\hat{X}\| = \sqrt{\lambda_{\text{max}}(\hat{X}^\dagger \hat{X})})
$
but other norms could be used equally well. For instance, see Appendix \ref{AppendixF} for results obtained for the Frobenius norm of the same quantity.

This measure quantifies the distance between a model Hamiltonian $\hat{H}_{\mathrm{nh}}$ and its Hermitian conjugate. It naturally vanishes for Hermitian systems.  
We find it most useful to normalize $\mathcal{D}$ to the norm 
of $\hat{H}_{\mathrm{nh}}$, in order to avoid attributing significant non-Hermiticity to systems that are actually dominated by a large Hermitian part (see also Appendix~\ref{appendixC} for a discussion about the unnormalized version). 
While it captures properties that the entire model system can in principle assume, quantifying non-Hermiticity via $\mathcal{D}$ requires access to the full Hamiltonian matrix.

\subsubsection{Non-Hermiticity score of observables}
\label{sec:score_observables}
For practical purposes (in numerics or experiment), it may be of interest to construct quantifiers that---in contrast to $\mathcal{D}$---rely only on the use of observables. 
To this end, we introduce the notion of ``non-Hermiticity score" of an observable, defined as 
\begin{equation}
{SC}^\mathcal{F}_{\mathrm{nh}}=|\mathcal{F}_{RR}[\rho_{RR}]-\mathcal{F}_{LL}[\rho_{LL}]|\,.
\label{eqn:NH_score}
\end{equation}
This score quantifies the  non-Hermiticity at the level of any computable physical quantity related to the model. It measures the difference of a state-dependent function   $\mathcal{F}[.]$  when evaluated for the state evolved under $\hat{H}_{\mathrm{nh}}$ or under $\hat{H}_{\mathrm{nh}}^\dagger$, which as discussed above lead to ensembles of right or left eigenvectors ($\rho_{RR}(t)$ or $\rho_{LL}(t)$), respectively. 
As illustrated in the next section, one can choose for $\mathcal{F}[.]$ single-body observables such as local magnetization or occupations as well as many-body observables such as von-Neumann entropy or purity. 
Following the discussions in Sec.~\ref{sec:kinematic}, the corresponding expectation values are well-defined. 
The non-Hermiticity score thus expresses how much the non-Hermiticity of the model Hamiltonian is concretely translated into physical observables in a given time evolution. 

Following also the discussion in Appendix~\ref{sec:eigenbasis}, the density matrices $\rho_{RR}$ and $\rho_{LL}$ used to evaluate the score can represent the snapshot of a system at a given evolution time (see also Fig.~\ref{fig:purity_score} below) or they can correspond to individual eigenstates (Figs.~\ref{fig:ent_score}). 
For the latter case, sometimes the non-Hermiticity score can depend significantly on the choice of eigenvectors. To have a analysis that captures the behavior across the spectrum, we can reformulate the non-Hermiticity score of Eq.~(\ref{eqn:NH_score}) as a well-defined distance measure, given by
\begin{eqnarray}  ||\overline{{SC}^\mathcal{F}_{\mathrm{nh}}}||_p=\Big( \sum_{i=1}\big({{SC}^\mathcal{F}_{\mathrm{nh}}}[i]\big)^{p} \Big)^{\frac{1}{p}}.
\end{eqnarray}
This quantity is nothing but the $p$-norm of the vector $\overline{{SC}^\mathcal{F}_{\mathrm{nh}}}$ with elements $\overline{{SC}^\mathcal{F}_{\mathrm{nh}}}=\{ {SC}^\mathcal{F}_{\mathrm{nh}}[1], {SC}^\mathcal{F}_{\mathrm{nh}}[2], \dots, {SC}^\mathcal{F}_{\mathrm{nh}}[D]\}$, where ${SC}^\mathcal{F}_{\mathrm{nh}}[k]$ denotes the non-Hermiticity score computed for the $k$-th eigenstate of the model, and $D$ corresponds to the Hilbert space dimension. 
In our case, we will illustrate the results obtained for  the case $p \rightarrow \infty$ (see Fig.~\ref{fig:Hatano_nelson}), for which the above definition reduces to the infinity norm, $||\overline{{SC}^\mathcal{F}_{\mathrm{nh}}}||_{\infty}=\max_i {SC}^\mathcal{F}_{\mathrm{nh}}[i]$, which is the maximum value of the non-Hermiticity score computed over all the eigenstates of the model. 

At the opposite limit, $p \rightarrow 0^+$, one obtains the $0$-norm, which counts the number of eigenvectors with non-zero non-Hermiticity score and thus captures the qualitative behavior across the entire spectrum. In our analysis below, for convenience we use a numerical threshold $\mathcal{E}_{\mathrm Th}^{\mathcal{F}}$, counting the total number of elements of $\overline{{SC}^\mathcal{F}_{\mathrm{nh}}}$ with the condition ${SC}^\mathcal{F}_{\mathrm{nh}}[k]\geq \mathcal{E}_{\mathrm Th}^{\mathcal{F}}$. We denote the corresponding score by $\mathcal{G}^{\mathcal{F}}$. An example is plotted in Fig.~\ref{fig:score_count}.

\section{Illustration at model systems}
\label{sec:models}
We are now equipped with ways to quantify the non-Hermiticity of a quantum system. In this section, we illustrate their behavior at two models. The first example has a quantum information-theoretic origin, namely, a non-Hermitian Hamiltonian constructed using a set of imperfect Bell states. In the second example, we consider a many-body Hamiltonian, the interacting Hatano--Nelson model, where non-Hermiticity is incorporated via asymmetric hopping between the sites. The findings related to both  models highlight the inequivalency between different notions of non-Hermiticity. 

\subsection{Non-Hermitian model construed using imperfect Bell state}
\label{subsec:bell}
In order to introduce the first example of the  non-Hermitian Hamiltonian that we have considered in our work, we present a set of ``imperfect'' Bell states characterized by the parameter $\alpha$ ($0< \alpha \leq 1$) and having the following mathematical form  when expressed in the computational basis, given by
\begin{eqnarray}
|R_{1,2}\rangle&=&\frac{(\frac{1}{\alpha}-1)|00\rangle + (|00\rangle\pm|11\rangle)}{\sqrt{1+\frac{1}{\alpha^2}}}\,, \nonumber\\
|R_{3,4}\rangle&=&\frac{(1-\alpha)|00\rangle+ (|01\rangle \pm|10\rangle)}{\sqrt{(1-\alpha)^2+2}}\,.
\label{eqn:right_vec}
\end{eqnarray}
For $\alpha=1$, the above set of states reduces to the perfect set of Bell states, $|\Psi^{\pm}\rangle=\frac{|00\rangle \pm |11\rangle}{\sqrt{2}}$, and  $|\Phi^{\pm}\rangle=\frac{|01\rangle \pm |10\rangle}{\sqrt{2}}$. It can be prepared employing    Bell state measurements (BSM) that project two qubits onto one of the four maximally entangled Bell states, which has been efficiently realized in atomic \cite{BSM_atomic} and optical platforms \cite{BSM_optical}. Imperfect Bell states such as the above could be the result of preparation in a non-ideal BSM setup. 

The non-Hermitian Hamiltonian ($\hat{H}^{\mathrm{Bell}}_{\mathrm{nh}}$) is constructed by choosing the imperfect Bell states as the set of right eigenvectors. Subsequently, the set of left eigenvectors can be obtained from the inverse of a matrix whose columns are the vectors $\left\{| R_m\rangle\right\}$. The unnormalized left eigenvectors are
\begin{eqnarray}
|L_{1,2}\rangle&=&\frac{\sqrt{1+\alpha^2}}{2}(|00\rangle+(\alpha-1) |01\rangle\pm \frac{1}{\alpha}|11\rangle),\nonumber\\
|L_{3,4}\rangle&=&\frac{\sqrt{(1-\alpha)^2+2}}{2}(|01\rangle\pm|10\rangle)\,.\label{eqn:left_vec}
\end{eqnarray}
One can easily verify that the above vectors form  a biorthonormal basis satisfying the relation
\begin{equation}
\langle L_m|R_n\rangle=\delta_{mn}\,,\qquad \langle R_m|R_m\rangle=1.
\end{equation}
We now construct a non-Hermitian  Hamiltonian using the basis states described in Eqs.~(\ref{eqn:right_vec}) and (\ref{eqn:left_vec}) as
\begin{eqnarray}
\hat{H}^{\mathrm{Bell}}_{\mathrm{nh}}=\sum_m \lambda_m |R_m\rangle \langle L_m|\,, 
\label{eqn:Bell_ham}
\end{eqnarray}
where we choose the $\lambda_m$ as real numbers \cite{Blender}, sampled from independent Gaussian distributions with zero mean and unit variance \footnotemark[1]. \footnotetext[1]{We find that for other choices of the variance there is no significant change in the behavior of $\mathcal{D}_{\mathrm{min}(max)}$} 
In recent years, such a bottom-up approach of reverse engineering a non-Hermitian quantum Hamiltonian from  easily constructed eigenvectors got much attention in the community  \cite{nh_Hamiltonian_MPS}. Below, we present the results obtained   for both static and dynamical quantum properties related to the above non-Hermitian Hamiltonian. \\\\

\begin{figure}[t]
 \includegraphics[width=\linewidth]{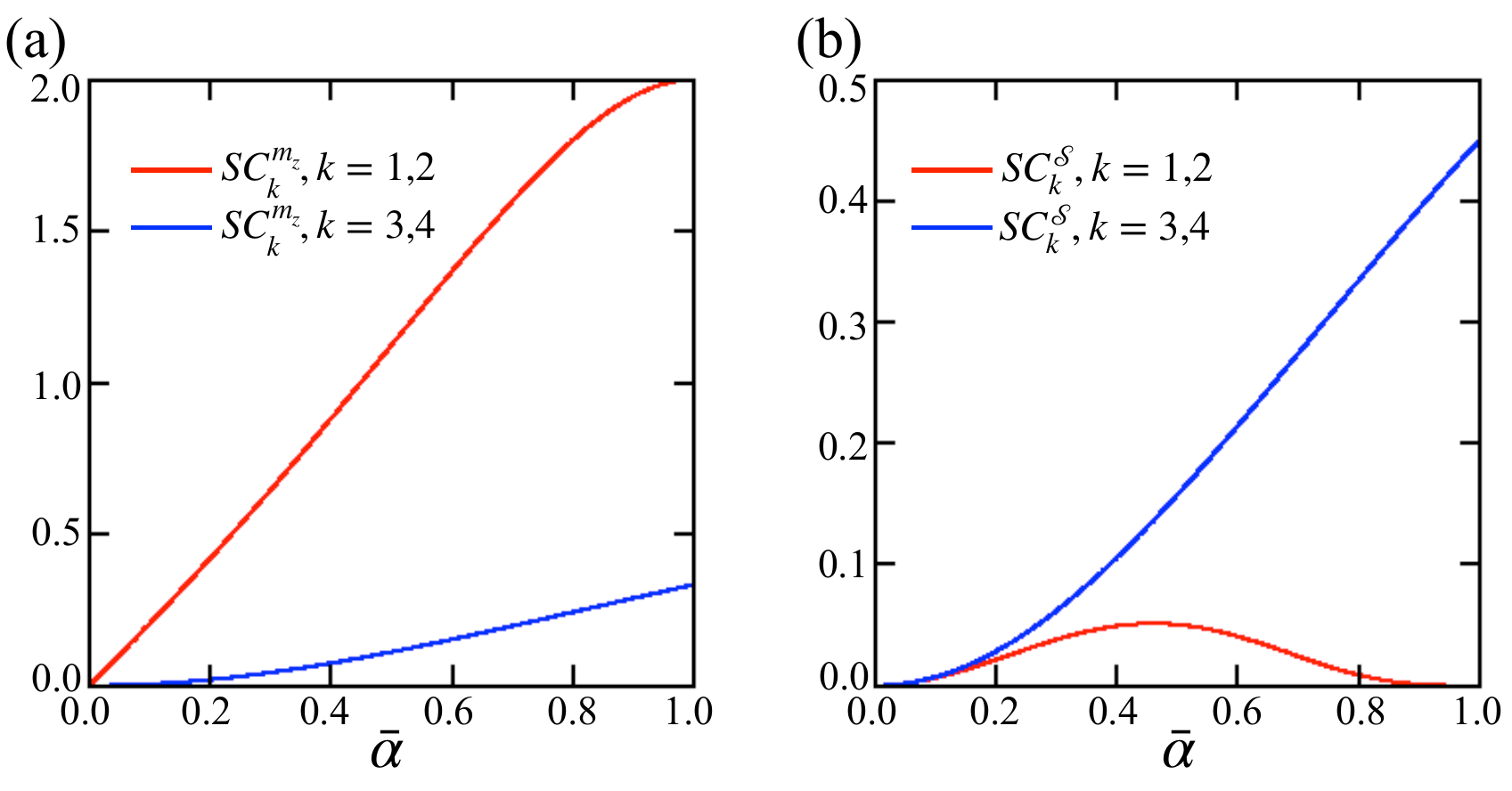}
 \caption{Non-Hermiticity at the level of single- and two-site quantum properties of the model constructed using imperfect Bell states. 
 (a) Non-Hermiticity score $SC^{m_z}$ of a single-site quantum property, the $z$-magnetization, computed for both right as well as left eigenvectors. 
 The maximum   and the minimum ($SC^{m_z}_k$) values of the score  are obtained for $k \in 1,2$ and $k\in 3,4$, respectively. In both cases, the behavior remains qualitatively the same as the Hamiltonian non-Hermiticity shown in Fig.~\ref{fig:Ham}(a). 
 (b) Entanglement score, $SC^{\mathcal{S}}$, obtained for the von Neumann entropy. Interestingly, the behavior obtained for quantum entanglement differs from the single-site score. In particular, unlike $SC^{m_z}$, $SC^{\mathcal{S}}_k$ attains its  minimum value  for $k=1,2$ and becomes maximum  for $k=3, 4$. Moreover,  though $SC^{\mathcal{S}}_{k=3,4}$  shows monotonic growth similar to $SC^{m_z}$ and $\mathcal{D}_{\min,\max}$, $SC^{S}_{k=1,2}$ vanishes for low as well as high values of $\bar{\alpha}$ and attains a maximum around $\bar{\alpha} = 0.449	$.  }
  \label{fig:ent_score}  
\end{figure}

\subsubsection{Non-Hermiticity at the level of  Hamiltonian: \\behavior of  $\mathcal{D}$ with $\alpha$}
We first examine how the degree of non-Hermiticity $\mathcal{D}$ behaves for the Hamiltonian described in Eq.~(\ref{eqn:Bell_ham}). For a fixed value of the parameter $\alpha$, we compute $\mathcal{D}$ for 1000 random realizations of $\hat{H}_{\mathrm{nh}}$ by choosing its eigenvalues $\{\lambda_m\}$ from a Gaussian distribution with zero mean and unit variance. In Fig.~\ref{fig:Ham}(a), we plot the maximum as well as minimum among all the random realizations,  denoted by $\mathcal{D}_{\max}$ and  $\mathcal{D}_{\min}$, respectively, against $\bar{\alpha}=1-\alpha$. At $\alpha=1$ ($\bar{\alpha}=0$), the right and left eigenvectors recover perfect Bell states and the Hamiltonian becomes Hermitian, reflected in vanishing $\mathcal{D}$. Though fluctuating, $\mathcal{D}_{\min}$ tends to increase with $\bar{\alpha}$. The growth of $\mathcal{D}_{\max}$ is smoother and at a faster rate, reaching the  maximum non-Hermiticity of $\mathcal{D}=1$ at $\bar{\alpha}=1$. 

\subsubsection{Non-Hermiticity score for single-site property: magnetization}
One of the main goals of our work is to examine to what extent the non-Hermiticity of the model Hamiltonian gets translated to its physical properties. Towards that aim, we now analyze the non-Hermiticity score for the magnetization in individual eigenstates. Results for quantum entanglement in individual eigenstates are presented in the following section. Analytical forms of all the quantities discussed in these two sections are provided  in Appendix~\ref{appendixA}. 
Two subsequent sections contain results for purity and entanglement under time evolution.

In Fig.~\ref{fig:ent_score}(a), we plot the non-Hermiticity score for the magnetization of the first qubit (labelled $A$) along the $z$-axis, $\mathcal{F}[\rho]=m_z[\rho]=\text{Tr}(\hat{\sigma}^z_A \rho)$, 
$SC^{m_z}_k=|m_z[\rho_{RR}^k]-m_z[\rho_{LL}^k]|$ 
Here, $\hat{\sigma}^z$ is the $2\times2$ Pauli matrix along the $z$-direction and $\rho_{RR}^k=\frac{|R_k\rangle\langle R_k|}{\langle R_k| R_k \rangle}$,  
$\rho_{LL}^k=\frac{|L_k\rangle\langle L_k|}{\langle L_k| L_k \rangle}$, where $|R_k\rangle$ and $|L_k\rangle$ are individual right and left eigenvectors of $H^{\mathrm{Bell}}_{\mathrm{nh}}$, as given in Eqs.~(\ref{eqn:right_vec}) and (\ref{eqn:left_vec}). 
As we are considering here properties of single eigenvectors, the eigenvalues $\lambda_k$ do not directly enter the non-Hermiticity score, in contrast to $\mathcal{D}$ discussed above, and no sampling over $\lambda_k$ is necessary. 
To give an operational interpretation of the above observables, we may want to assume the eigenvalues to fulfill case 2 in Sec.~\ref{sec:eigenbasis}, such that $\rho_{RR}^k$  and $\rho_{LL}^k$ are prepared when evolving an initial state to long times under $\hat{H}^{\mathrm{Bell}}_{\mathrm{nh}}$ and $(\hat{H}^{\mathrm{Bell}}_{\mathrm{nh}})^\dagger$, respectively. 

As Fig.~\ref{fig:ent_score}(a) shows, at each value of $\bar{\alpha}$,  $SC^{m_z}_k$ achieves its maximum (minimum) for $k\in 1, 2$ ($3, 4$). Both the minimum and maximum of $SC^{m_z}$ increase monotonically with $\bar{\alpha}$ and agree qualitatively with $\mathcal{D}$ as shown in Fig.~\ref{fig:Ham}(a). 
Moreover, $SC^{m_z}_{k=1,2}$ attains its theoretical maximum at $\bar{\alpha}=1$. 
In other words, this single-particle property can serve as a probe of the non-Hermiticity of the model. Notably, for all values of $\alpha$ and any eigenvector $k$, $m_z$ remains zero when computed for $\rho_{RL}^k=|R_k\rangle\langle L_k|$. Hence, the single-site quantum property obtained in the biorthogonal basis fails to capture any signature of the non-Hermiticity of the model.  

\subsubsection{Non-Hermiticity score  for the von Neumann entropy (static case)}

In this section, we analyze entanglement properties of the same set of right and left eigenvectors as considered above, in particular of the von Neumann entropy (VNE), defined as $\mathcal{S}[\hat{\rho}]=-\text{Tr}_A(\text{Tr}_B(\hat{\rho}) \log_2 \text{Tr}_B(\hat{\rho}))$. 
Figure~\ref{fig:ent_score}(b) presents the associated non-Hermiticity score, $SC^{\mathcal{\mathcal{S}}}_k=|\mathcal{S}[\hat{\rho}_{RR}^k]-\mathcal{S}[\hat{\rho}_{LL}^k]|$. 
Unlike the case for $m_z$, at each value of  $\bar{\alpha}$, $SC^{\mathcal{S}}_k$ becomes minimum for $k=1, 2$,  and the behavior does not comply with $\mathcal{D}$.  In particular, $SC^{\mathcal{S}}_{k=1,2}$ vanishes both for  $\bar{\alpha} \rightarrow 0$ and  $\bar{\alpha} \rightarrow 1$, and attains a maximum around $\bar{\alpha} = 0.449$. 
This is significantly different from the behavior of $\mathcal{D}_{\max}$ (or $\mathcal{D}_{\min}$) as well as the single-site property discussed above. 

In contrast, the maximum value of $SC^{\mathcal{S}}_k$, attained for all values of $\bar{\alpha}$ at $k=3, 4$, remains  akin to $\mathcal{D}$. Nevertheless, quantitative differences remain. E.g., unlike $\mathcal{D}$, $SC^{\mathcal{S}}_{k=3,4}$ increases almost linearly in the region  $0.4 \lesssim \bar{\alpha} \leq 1.0$ and goes to zero at $\bar{\alpha}=0$. 
Moreover, $SC^{\mathcal{S}}_{k=3,4}$ does not reach its theoretical maximum of $1$ at $\bar{\alpha}=1$. 
Again, the VNE obtained for the biorthogonal ensemble $\hat{\rho}_{RL}^k=|R_k\rangle \langle L_k|$ remains independent of $\bar{\alpha}$ (yielding the value $\mathcal{S}(|R_k\rangle \langle L_k|)=\frac{1}{2}$  $\forall k$) and is thus not a useful probe for the non-Hermiticity of the system.

\subsubsection{Non-Hermiticity score of global purity} 
\begin{figure}[t!]
  \includegraphics[width=\linewidth]{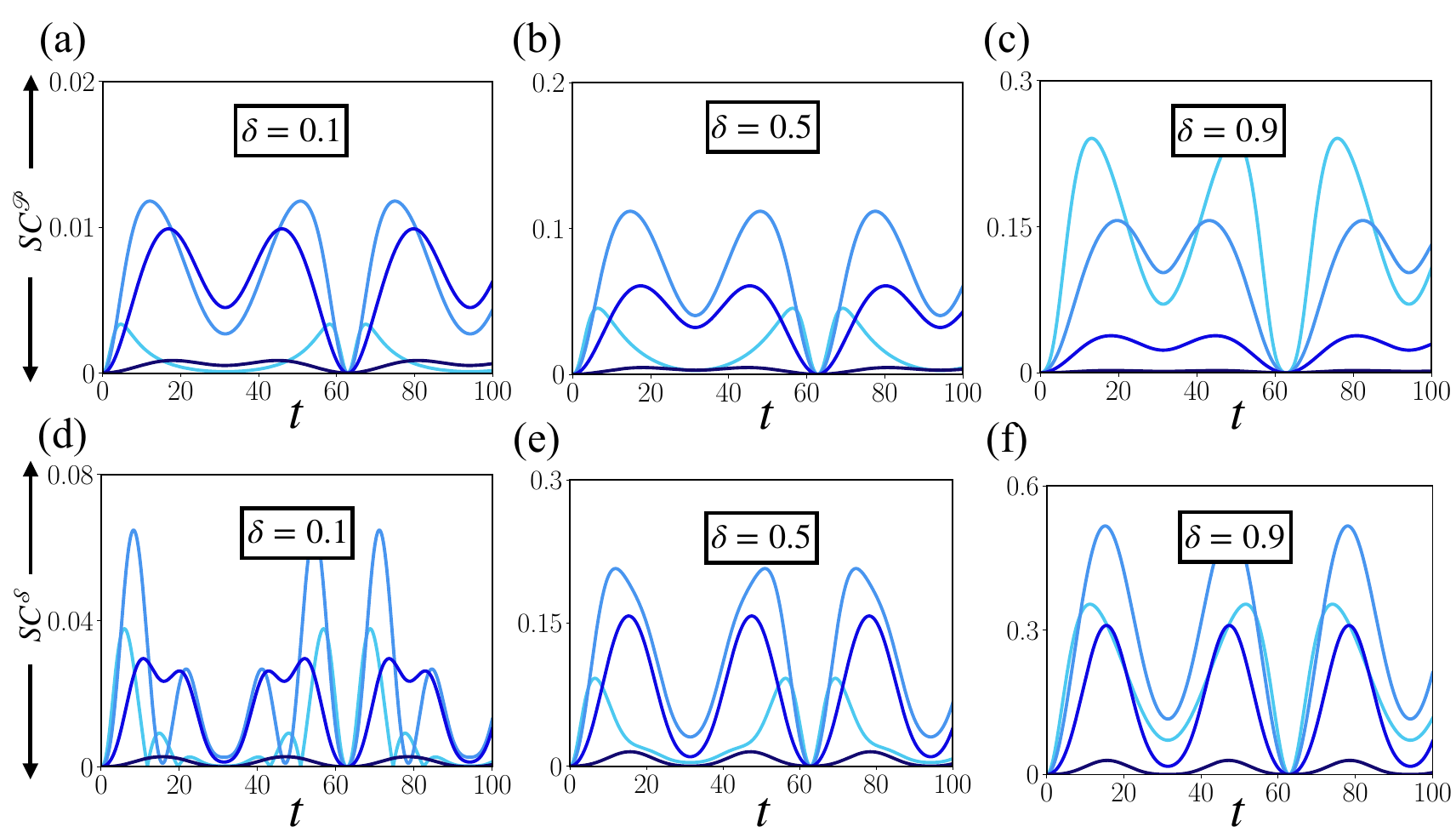}
 \caption{Behavior of non-Hermiticity score with time,  for global purity $\mathcal{P}$ and von Neumann entropy $\mathcal{S}$, computed for an initial state $\hat{\rho}_W=\delta |\Psi^-\rangle \langle \Psi^-|+ (1-\delta)\frac{\mathbb{I}}{4}$ subjected to evolution under $\hat{H}^{\mathrm{Bell}}_{\mathrm{nh}}$ and  $(\hat{H}^{\mathrm{Bell}}_{\mathrm{nh}})^\dagger$. 
 We probe the cases for $\alpha$=0.1, 0.3, 0.6, 0.9 as shown using curves with darker to lighter shades of blue, and in all cases fix the Hamiltonian eigenvalues to $\lambda_1=0.1,\ \lambda_2=0.2, \ \lambda_3=0.3, \ \lambda_4=0.4$. 
 (a)-(c) Purity score, $SC^{\mathcal{P}}(t)$ for (a) $\delta=0.1$, (b) $\delta=0.5$, and (c) $\delta=0.9$. For large $\delta$, the amplitude of $SC^{\mathcal{P}}(t)$ decreases with increasing $\alpha$, while for small $\delta$ it attains a maximum at intermediate values of $\alpha$.  
 (d)-(f) Non-Hermiticity score  for the VNE, $SC^{\mathcal{S}}(t)$, for same choices of $\delta$.}
  \label{fig:purity_score}  
\end{figure}

In this section, we study the time-dependent non-Hermiticity score of the purity, $\mathcal{P}[\hat{\rho}(t)]={\text{Tr}\Big(\hat{\rho}(t)^2\Big)}/{\text{Tr}(\hat{\rho}(t))^2}$, of a state $\hat{\rho}(t)$ obtained from an initial state $\hat{\rho}_{\mathrm{in}}$  when evolved under $\hat{H}^{\mathrm{Bell}}_{\mathrm{nh}}$ or $(\hat{H}^{\mathrm{Bell}}_{\mathrm{nh}})^\dagger$. 
If the initial state is chosen as maximally mixed, $\hat{\rho}_{\mathrm{in}}=\frac{\mathbb{I}}{4}$, one gets $\mathcal{P}(\hat{\rho}_{RR}(t))=\mathcal{P}(\hat{\rho}_{LL}(t))$ (see Appendix~\ref{appendixB}). Hence,  $SC^{\mathcal{P}}(t)=0$ at all times and  fails to witness any imprint of $\mathcal{D}$. 
However, for a non-maximally mixed initial state, $SC^{\mathcal{P}}(t)$ acquires a non-zero value. In Fig.~\ref{fig:purity_score},  we plot the time-dependent purity score for systems initialized in a mixed Werner state, $\hat{\rho}_W=\delta |\Psi^-\rangle \langle \Psi^-|+ (1-\delta)\frac{\mathbb{I}}{4}$, with $|\Psi^-\rangle=\frac{|01\rangle-|10\rangle}{\sqrt{2}}$. We tune the initial purity to (a) $\delta=0.1$, (b) $\delta=0.5$, and (c) $\delta=0.9$, and choose four different values of $\alpha$, $\alpha= 0.1$, $0.3$, $0.6$, $0.9$, which are depicted using lighter to darker shades of blue. In these plots, we fix the Hamiltonian eigenvalues to $\lambda_1=0.1$, $\lambda_2=0.2$, $\lambda_3=0.3$, and $\lambda_4=0.4$. 

For all values of $\alpha$ and $\delta$, $SC^{\mathcal{P}}(t)$ exhibits non-monotonic, periodic behavior with time. Moreover, at low $\delta$, the highest value of the non-Hermiticity score within the considered time window,  $SC^{\mathcal{P}}_{\max}=\max_{t\in[0,100]}\{SC^{\mathcal{P}}(t)\}$,   depends on the parameter $\alpha$. For instance,  both at low and high values of $\alpha$,  $SC^{\mathcal{P}}_{\max}$ remains low while it attains larger values at intermediate $\alpha$. 
However, for initial states with higher purity, i.e., with larger $\delta$ [exemplified for $\delta=0.9$ in Fig.~\ref{fig:purity_score}(c)] $SC^{\mathcal{P}}_{\max}$ decreases with increasing $\alpha$.  All these observations suggest $\mathcal{P}$ does not imitate the profile of $\mathcal{D}$, and thus gives an independent imprint of non-Hermiticity on the time evolution.

\subsubsection{Non-Hermiticity score of von Neumann entropy (dynamic case)}
Further above, we have discussed the non-Hermiticity score for the VNE when computed for the special cases of $\hat{\rho}_{RR}(t)\overset{ t=\infty}{\rightarrow}\hat{\rho}_{RR}^k=|R_k\rangle \langle R_k|$ and $\hat{\rho}_{LL}(t)\overset{ t=\infty}{\rightarrow}\hat{\rho}_{LL}^k=|L_k\rangle \langle L_k|$. In this section, we consider a more general scenario where  $\hat{\rho}_{RR}(t)$  is obtained following Eq.~(\ref{eqn:dynamics1}) (and mutatis mutandis for $\hat{\rho}_{LL}(t)$). 
Note that, in this case, the VNE is not an entanglement measure, as the global states $\hat{\rho}_{RR}(t)$ and $\hat{\rho}_{LL}(t)$ are not pure. 
In Fig.~\ref{fig:purity_score}(d)-(f), we plot the time-dependent behavior of $SC^{\mathcal{S}}(t)=|\mathcal{S}[\hat{\rho}_{RR}(t)]-\mathcal{S}[\hat{\rho}_{LL}(t)]|$, for the same parameter values as in panels (a)-(c). Similar to the purity score, $SC^{\mathcal{S}}$ exhibits non-monotonic behavior with time and its maximum value computed for the considered time period, $SC^{\mathcal{S}}_{\max}=\max_{t\in[0,100]}\{SC^{\mathcal{S}}(t)\}$, again remains low for high values of $\alpha$. However, this score achieves significant values at small $\alpha$ even at large $\delta$.   

As these examples show, the non-Hermiticity score for dynamical physical quantities does not necessarily follow the profile of $\mathcal{D}$. In particular, $SC^{\mathcal{F}}$ can attain a  low value even when the model Hamiltonian is highly non-Hermitian (in this case $\alpha \rightarrow 0$). 
One must thus distinguish between the non-Hermiticity of the model Hamiltonian and its imprint onto physical properties.

\subsection{Non-Hermitian interacting Hatano--Nelson model}
\label{subsec:hn}

The second example system we consider is an interacting quantum many-body Hamiltonian, the non-Hermitian Hatano--Nelson model \cite{Hatano1,Hatano2,Hatano3,HN_interacting}, defined by 
\begin{equation}\label{eq:hatano_nelson}  \hat{H}^{\mathrm{HN}}_{\mathrm{nh}}=\sum_i^N -J( e^\chi \hat{c}^\dagger_i \hat{c}_{i+1}+e^{-\chi} \hat{c}^\dagger_{i+1} \hat{c}_{i})+V \hat{n}_i \hat{n}_{i+1}\,.
\end{equation}
Here, the hopping strength $J=1$ scales the Hamiltonian, $\hat{c}_i^\dagger$($\hat{c}_i$) is the creation (annihilation) operator for a spinless fermionic particle at site $i$, $\hat{n}_i=\hat{c}_i^\dagger \hat{c}_i$ the single-mode number operator, and $V$ is the strength of nearest-neighbor interaction. Moreover, 
we restrict our studies to the half-filling sector and employ periodic (anti-periodic) boundary conditions for 
odd (even) $N/2$.
The notion of non-Hermiticity is introduced by asymmetric hopping controlled by the parameter $\chi$, where the model reduces to the Hermitian version at $\chi=0$. 
After explaining in Appendix~\ref{sec:kinematic} how non-Hermitian dynamics can be understood operationally as a specific case of open quantum system dynamics, we present in Appendix~\ref{sec:hatanorealization} a detailed discussion on how the dynamics of the Hatano-Nelson Hamiltonian—and its Hermitian conjugate—can be implemented in modern cold atom quantum systems via open quantum system techniques.
The model has parity-time (PT) reversal symmetry, which results in eigen energies appearing as complex conjugate pairs. In Ref.~\cite{HN_interacting}, it is reported that as the strength of the interaction ($V$) increases,  the model undergoes two PT symmetry-breaking transitions: One of them is due to an exceptional point between the two lowest excited states and marked by the non-analytic behavior of the finite-size level-spacing $\Delta_{01}:=\mathrm{Re}(\lambda_1 - \lambda_0)$. The other corresponds to a full collapse of the complex many-body spectrum onto the real axis in a finite-size system. The former transition is accompanied by a first-order symmetry-breaking transition from a PT broken gapless phase to a PT-symmetric charge-density phase. Similar characteristics are also reported in the hardcore bosonic version of the model with periodic boundary condition \cite{HN_hc_boson}.  
In what follows, we present the results for the degree of Hamiltonian non-Hermiticity of the model as well as single- and multi-site non-Hermiticity scores, and also contrast their trend against the above-mentioned transitions.

\subsubsection{Hamiltonian non-Hermiticity ($\mathcal{D}$) in the interplay of asymmetric coupling ($\chi$) and interactions ($V$)}
To begin with, we examine how the asymmetric coupling parameterized by $\chi$ and interaction strength $V$ affect the degree of non-Hermiticity ($\mathcal{D}$) of the Hamiltonian $\hat{H}^{\mathrm{HN}}_{\mathrm{nh}}$, plotted in Fig.~\ref{fig:Ham}(b). Note that here, and in all subsequent plots, the 
parameter $V$ is presented on a logarithmic scale. At $V=0$, $\mathcal{D}$ grows monotonously with $\chi$ and eventually saturates to the theoretical maximum of $\mathcal{D}\approx 2$.  However, turning on even a small amount of interaction has a detrimental effect on $\mathcal{D}$: it attains a significant non-zero value only beyond a non-vanishing threshold coupling, $\chi_c$. Interestingly, we further observe that up to $V\approx 10^3$, $\chi_c$ increases almost linearly with $V$. At even larger interaction strengths, $\mathcal{D}$ remains zero up to the largest values of $\chi$ considered in our analysis ($\chi= 5.4$). 

As these results show, interactions suppress the non-Hermiticity of the model significantly, as the distance between $\hat{H}_{\mathrm{nh}}^{\mathrm{HN}}$ and $(\hat{H}_{\mathrm{nh}}^{\mathrm{HN}})^\dagger$ becomes insignificant in comparison to the norm of $\hat{H}_{\mathrm{nh}}^{\mathrm{HN}}$. Interestingly, the domain of vanishing $\mathcal{D}$ remains close to the region of non-analyticity of $\Delta_{01}$, which can be read off from Fig.~\ref{fig:Ham}(b) and its inset. This indicates that $\mathcal{D}$ can be useful in qualitatively detecting the PT symmetry-breaking transition present in this model. Furthermore, in Appendix \ref{AppendixF}, we report  that the Frobenius norm exhibits qualitatively similar behavior. However, it is important to note that whether this behavior holds for any generic non-Hermitian Hamiltonian remains an open question and warrants further investigation. In turn, this may signify the importance of the normalization of $\mathcal{D}$ considered in Eq.~\eqref{eq:Hnhscore}. In fact, the unnormalized incarnation of $\mathcal{D}$ does not reveal any transition (see Appendix \ref{appendixC}). 
In the following, we will examine to what extent such a behavior reflects in the behavior of single- and multi-site quantum properties related to the model. 

\begin{figure}
  \includegraphics[width=\linewidth]{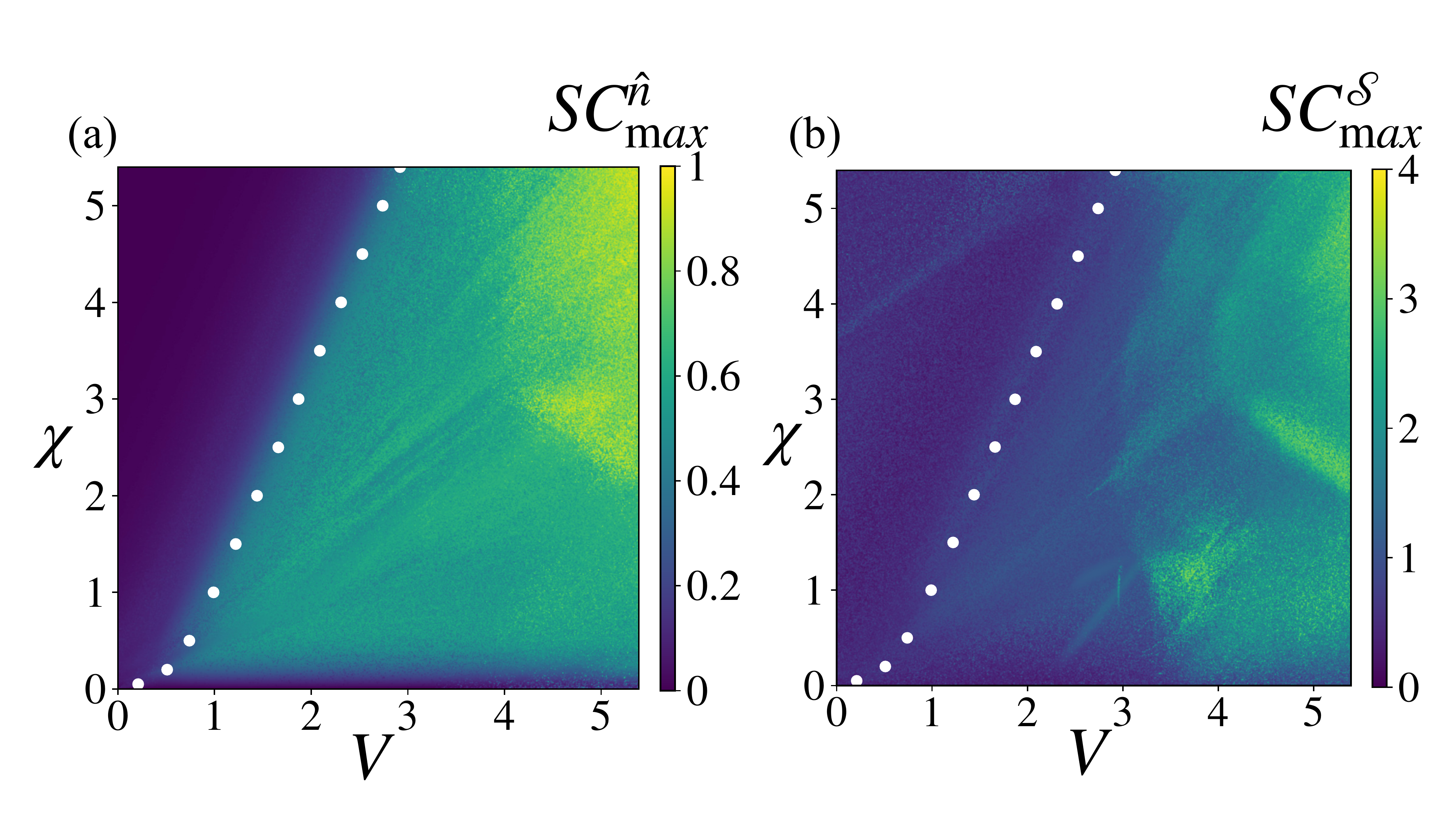}
 \caption{Comparison of non-Hermiticity at the level of the Hamiltonian and single- and multi-party physical properties of the non-Hermitian Hatano--Nelson model ($\hat{H}_{\mathrm{nh}}^{\mathrm{HN}}$). We compute the non-Hermiticty score for both the number operator $\hat{n}$ and half-chain VNE $\mathcal{S}$ for all the right and left eigenvectors of the model and finally compute the maximum among all of them. (a) Depicts the  behavior of $SC^{\hat{n}}_{\max}$ in the $V-\chi$ plane. We note that for a large region in the parameter space, the behavior of $SC^{\hat{n}}_{\max}$ remains almost complementary to $\mathcal{D}$, as shown in  Fig.~\ref{fig:Ham}(b). A similar plot for the half-chain VNE is presented in (b), where for the same set of parameters, we plot the behavior of $SC^{\mathcal{S}}_{\max}$. We note that $SC^{\mathcal{S}}_{\max}$ behaves differently from both $SC^{\hat{n}}_{\max}$ as well as the degree of non-Hermiticity ($\mathcal{D}$).  In particular, $SC^{\mathcal{S}}_{\max}$ attains significant value only when both $V$ and $\chi$ take large values. However, closer inspection shows the qualitative agreement between the trend of $SC^{\hat{n}}_{\max}$ and $SC^{\mathcal{S}}_{\max}$.
 The white circles correspond to the PT symmetry-breaking transition marked by the non-analytic behavior of $\Delta_{01}$ defined in Sec. \ref{subsec:hn} and shown in the inset of Fig. \ref{fig:Ham}. We notice the qualitative agreement between the PT symmetry-breaking transition and domain for vanishing scores. The data is reported for $N = 12$ sites, half-filling, and with anti-periodic boundary conditions. 
  }
  \label{fig:Hatano_nelson}  
\end{figure}

\subsubsection{Single-site properties: non-Hermiticity score of number operator $\hat{n}$}

\begin{figure}
  \includegraphics[width=\linewidth]{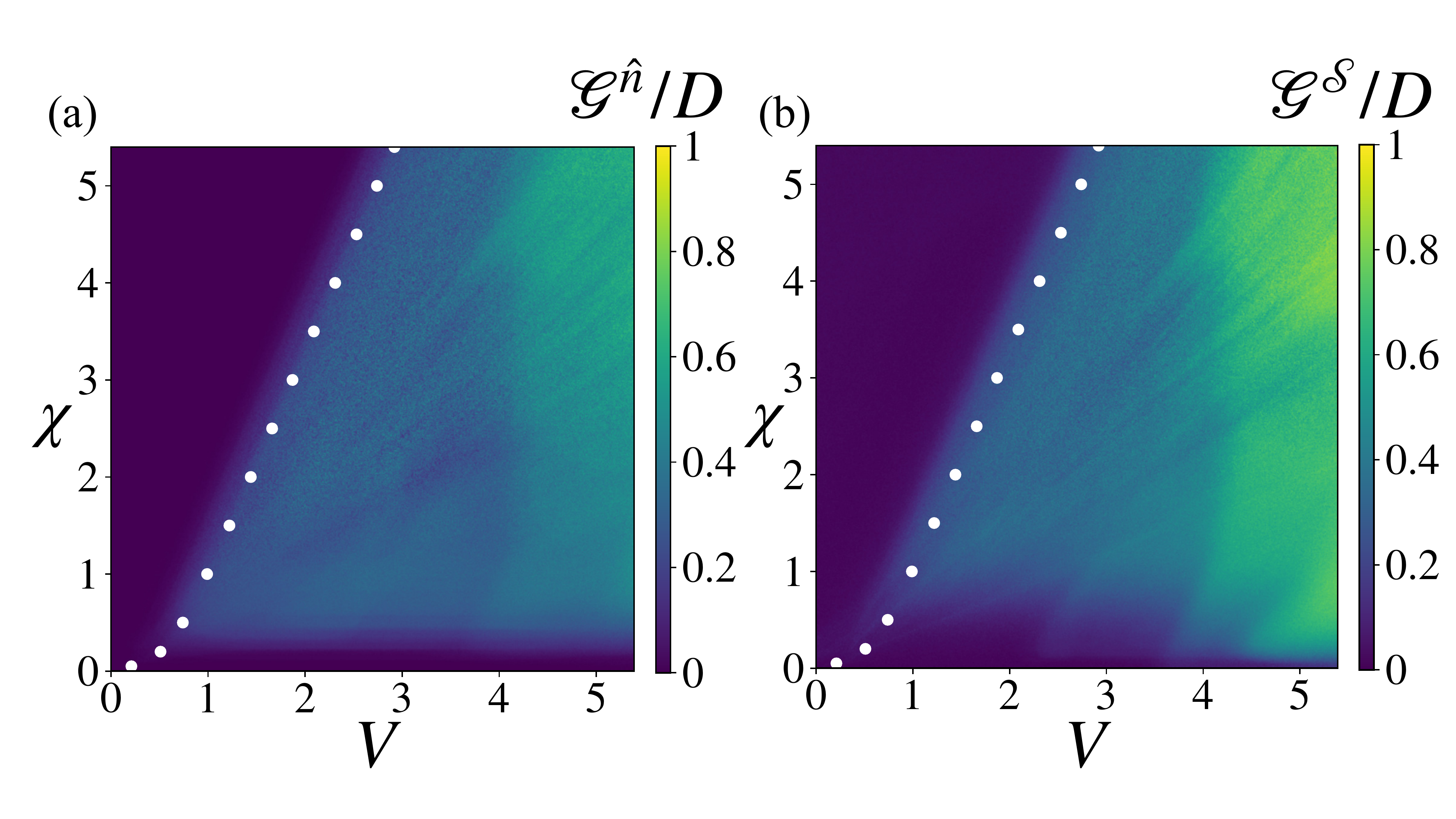}
 \caption{Global characteristics of $\overline{{SC}^\mathcal{F}_{\mathrm{nh}}}$ for the non-Hermitian Hatano--Nelson model ($H_{\mathrm{nh}}^{\mathrm{HN}}$). We compute the total number of elements of $\overline{{SC}^{\hat{n}}_{\mathrm{nh}}}$ and $\overline{{SC}^{\mathcal{S}}_{\mathrm{nh}}}$ with the conditions ${SC}^{\hat{n}}_{\mathrm{nh}}[k]\geq \mathcal{E}_{\mathrm{Th}}^{\hat{n}}$ and ${SC}^{\mathcal{S}}_{\mathrm{nh}}[k]\geq \mathcal{E}_{\mathrm{Th}}^{\mathcal{S}}$, respectively, where $k = 1,2,...,D$ with $D$ being the Hilbert space dimension. These quantities, scaled by $D$, are shown in (a) and (b). We obtain a similar trend as in Fig. \ref{fig:Hatano_nelson}. Data for threshold values $\mathcal{E}_{\mathrm Th}^{\hat{n}} = \mathcal{E}_{\mathrm Th}^{\mathcal{S}} = 0.1$. Similar to Figs. \ref{fig:Ham} and \ref{fig:Hatano_nelson}, the white circles correspond to the PT symmetry-breaking transition. The data is reported for $N = 12$ sites, half-filling, and with anti-periodic boundary conditions. 
 }
  \label{fig:score_count}  
\end{figure}

To study the impact of non-Hermiticity on single-site properties, we analyze the non-Hermiticity score of the number operator for the first site, $\hat{n}_1$.
The maximum score over all eigenstates, $SC^{\hat{n}}_{\max}\equiv ||\overline{{SC}^{\hat{n}}_{\mathrm{nh}}}||_{\infty}$, is plotted in  Fig.~\ref{fig:Hatano_nelson}(a). Importantly, $SC^{\hat{n}}_{\max}$ exhibits an almost complementary behavior to the non-Hermiticity of the Hamiltonian $\mathcal{D}$.  In particular, at low values of interaction, $SC^{\hat{n}}_{\max}$ becomes almost zero. Once the strength of the interaction is increased, expectation values of the single-site occupation evaluated in right and left eigenvectors start to exhibit a disparity, which is captured by significantly non-zero values of  $SC^{\hat{n}}_{\max}$ within an intermediate regime of $\chi$. Throughout the entire window of interaction strengths considered, $SC^{\hat{n}}$ vanishes at small $\chi$, while beyond $V\approx 10^2$ the non-vanishing region extends up to the maximum value of $\chi$ that we have considered. 
The region of $V$ where $SC^{\hat{n}}_{\max}$ becomes finite for any $\chi$ resides in close proximity to the PT symmetry-breaking transition marked by the non-analytic point of $\Delta_{01}$. Thereby, $SC^{\hat{n}}_{\max}$ can be considered as another candidate in identifying the PT symmetry-breaking transition of the model. 

In summary, the complementary behavior of  $SC^{\hat{n}}_{\max}$ to $\mathcal{D}$ suggests that the single-particle quantum property of the model can be a useful probe in detecting the non-Hermiticity of the model.  

Unlike the case of $\hat{H}_{\mathrm{nh}}^{\mathrm{Bell}}$, here the spectrum of the Hamiltonian consists of a large number of eigenvectors. Therefore, to illustrate the non-Hermitian behavior across the spectrum of the model, in Fig.~\ref{fig:score_count}(a) we plot the quantity $\mathcal{G}^{\hat{n}}$, see Sec.~\ref{sec:score_observables}, choosing as threshold value $\mathcal{E}_{\mathrm Th}^{\hat{n}} = 0.1$. The behavior is qualitatively similar to the one of $SC^{\hat{n}}_{\max}$, showing that in this case, the maximal score is representative of the behavior across the spectrum.

\subsubsection{Multi-site properties: non-Hermiticity score of von Neumann entropy}
As a multipartite quantum property, here we analyze the non-Hermiticity score computed for half-chain VNE, defined as $SC^{\mathcal{S}}=|\mathcal{S}[\rho_{RR}]-\mathcal{S}[\rho_{LL}]|$, where $\rho_{RR}=\text{Tr}_{N/2+1\dots N}(|R_k\rangle \langle R_k|)$ and $\rho_{LL}=\text{Tr}_{N/2+1\dots N}(|L_k\rangle \langle L_k|)$, with $|R_k\rangle$ and $\langle L_k|$ the $k$-th right and left vectors of $H^{\mathrm{HN}}_{\mathrm{nh}}$, respectively. In Fig.~\ref{fig:Hatano_nelson}(b), we plot the maximum of  $SC^{\mathcal{S}}$ over all the eigenstates of the model, $SC^{\mathcal{S}}_{\max}$. 

In this case, the behavior of $SC^{\mathcal{S}}_{\max}$ remains qualitatively similar to  $SC^{\hat{n}}_{\max}$.
In particular, $SC^{\mathcal{S}}_{\max}$  starts attaining significant nonzero value after a threshold value of the interaction strength that exhibits almost linear behavior with the coupling $\chi$.  Moreover, similar to the single-particle case, in most of the considered regimes, multipartite quantum properties of the right and left ensembles remain very much distinguishable from each other. Such a disparity becomes maximum when both interaction and the coupling become very large.

Furthermore, we report that  $\mathcal{G}^{\mathcal{S}}$, i.e., the number of eigenstates with $SC^{\mathcal{S}}$ above a threshold that we choose as $\mathcal{E}_{\mathrm Th}^{\mathcal{S}} = 0.1$, shows behavior similar to the $\mathcal{G}^{\hat{n}}$ as well as $SC^{\mathcal{S}}_{\max}$.
Similar to the previous cases, the PT symmetry-breaking transition region qualitatively coincides to the domain where $\mathcal{G}^{\mathcal{S}}$ becomes finite.

\section{Discussion}
\label{sec:conclusions}
In this work, we have introduced a formalism that quantifies different notions of non-Hermiticity, which can emerge in any generic non-Hermitian system. In particular, as first quantifier, we have introduced the Hamiltonian non-Hermiticity, a measure that computes the normalized distance between a non-Hermitian Hamiltonian $\hat{H}_{\mathrm{nh}}$ and its hermitian conjugate $(\hat{H}_{\mathrm{nh}})^\dagger$. 
We have introduced a second quantifier to capture a notion of non-Hermiticity at the level of physically computable quantities related to the model. 
This quantifier has been motivated by noticing that in the non-Hermitiam paradigm one can prepare quantum ensembles of right or left eigenvectors, which often yield incompatible results with respect to single- and multi-site quantum properties. 
Based on this observation, we have introduced the non-Hermiticity score that characterized the difference between observables in the right and left eigenensemble. 
These quantifiers enable a comparative study of the degree of non-Hermiticity across different regions of parameter space, for static as well as dynamic properties.

We have tested the formalism on two examples, in their static eigenstates as well as time evolution. The first non-Hermitian model we have considered has quantum information-theoretic origin, namely a non-Hermitian Hamiltonian constructed using a set of imperfect Bell states.  We showed that though the qualitative behavior of the non-Hermiticity score computed for a single-site observable complies with the degree of non-Hermiticity of the model, in the case of the von Neumann entropy their behavior significantly differs. 

Second, we have applied our formalism to a more generic case, to a quantum many-body model in the form of the interacting Hatano--Nelson model with asymmetric coupling. The model gives us a scope to examine how both interaction and hopping affect the behavior of distinct notions of non-Hermiticity. For the Hamiltonian non-Hermiticity,  interaction has an adverse effect. 
Importantly, for the non-Hermiticity score of single- and multi-site quantum properties, the behavior is more involved. For instance, the non-Hermiticity score of the single-site number operator $\hat{n}$ remains almost complementary to the behavior of the degree of non-Hermiticity of the Hamiltonian. This implies that even at large interactions, when the non-Hermitian part of the Hamiltonian gets strongly suppressed, locally the left and right vectors can be significantly different from each other. Finally, we have examined to what extent the non-Hermiticity of the model gets translated into the behavior of entanglement. At first glance it appears that the behavior of non-Hermiticity does not vary smoothly with both interaction strength and coupling. However, a careful look reveals that for a significant region of the considered parameter range, in relation to entanglement the left and right ensembles behave distinctly from each other and the difference qualitatively mimics the behavior obtained for the single-particle case. Furthermore, we obtain qualitative agreement between the PT symmetry-breaking transition regime of the model with the domain where the scores become zero to finite or vice-versa. This indicates that the non-Hermitian scores can be useful in identifying potential non-Hermitian phase transitions in a model.

In summary, in our work, we have opened a new route to examining the quantum properties of any generic non-Hermitian system. It permits to quantify non-Hermiticity, not only along the traditional lines that solely focus on the Hamiltonian of the model but also with respect to its influence onto observable properties. 
There exist already several instances of models in the literature for which it is known that the left and right eigenvectors show drastically different physical behavior, e.g., in terms of localization \cite{topology_nh2, Guo_2021} or of topological properties \cite{Ding_2022,Shen_2018}. From this knowledge, one can immediately construct relevant non-Hermiticity scores that witness the non-trivial non-Hermitian behavior of the model. In the future,  it will be interesting to apply our formalism to such and other physical systems to characterize exotic quantum many-body phases and identify potential non-Hermitian phase transitions.

An interesting complementary direction is to employ a bottom-up approach to construct non-Hermitian models with distinct entanglement properties---both bipartite and multipartite---as well as classical simulability features for their left and right eigenvectors. Such an approach could play a relevant role in preparing resourceful quantum states for advancing quantum technologies.

\acknowledgments{This project is funded by the European Union - Next Generation EU, Mission 4, Component 2 - CUP E53D23002240006.
This project has received funding from the Italian Ministry of University and Research (MUR) through the FARE grant for the project DAVNE (Grant R20PEX7Y3A). 
The project is funded under the National Recovery and Resilience Plan (NRRP), Mission 4 Component 2 Investment 1.4 - Call for tender No. 1031 of 17/06/2022 of Italian Ministry for University and Research funded by the European Union – NextGenerationEU (proj. nr. CN\_00000013).
This work was supported by the Swiss State Secretariat for Education, Research and innovation (SERI) under contract number UeMO19-5.1. 
Project DYNAMITE QUANTERA2\_00056 funded by the Ministry of University and Research through the ERANET COFUND QuantERA II – 2021 call and co-funded by the European Union (H2020, GA No 101017733). This work was supported   by the Caritro Foundation, the Provincia Autonoma di Trento, and Q@TN, the joint lab between University of Trento, FBK—Fondazione Bruno Kessler, INFN—National Institute for Nuclear Physics, and CNR—National Research Council.
S.B.\ acknowledges CINECA for the use of HPC resources under Italian SuperComputing Resource Allocation– ISCRA Class C Projects No. DISYK-HP10CGNZG9 and DeepSYK - HP10CAD1L3. S.S.R. acknowledges the faculty research scheme at IIT (ISM) Dhanbad, India, under Project No. FRS/2024/PHYSICS MISC0110. 

Views and opinions expressed are however those of the author(s) only and do not necessarily reflect those of the European Union or of the Ministry of University and Research. Neither the European Union nor the granting authority can be held responsible for them.
}

\appendix
\begin{widetext}

\section{Non-Hermiticity quantifiers using  other norms and observables}
\label{AppendixF}

This section provides additional results that reinforce our main conclusions. Beyond the operator norm employed in the definition of 
$\mathcal{D}$ in the main text, alternative norms--such as the Frobenius norm--can also be considered. When defined with respect to the Frobenius norm, we denote the Hamiltonian non-Hermiticity as $\mathcal{D}^{\prime}$. The numerator of $\mathcal{D}^{\prime}$ is then given by
${\mathcal{F}} = \sqrt{\mathrm{Tr}\Big(\left[\hat{H}^{\mathrm{HN}}_{\mathrm{nh}} - (\hat{H}^{\mathrm{HN}}_{\mathrm{nh}})^\dagger\right]^\dagger \left[\hat{H}^{\mathrm{HN}}_{\mathrm{nh}} - (\hat{H}^{\mathrm{HN}}_{\mathrm{nh}})^\dagger\right]\Big)}$. 
To compute it explicitly, we start from 
\begin{align}
\hat{H} - \hat{H}^\dagger = \sum_n \left( \varepsilon_n |R_n\rangle\langle L_n| - \varepsilon_n^* |L_n\rangle\langle R_n| \right),
\end{align}
and use the decompositions
\begin{align}
|R_n\rangle &= \sum_m \alpha_{nm}|R_m\rangle + \beta_{nm}|R_{n\perp m}\rangle, \qquad \alpha_{nm} = \langle R_m|R_n\rangle,\qquad\beta_{nm}=\langle R_{n \perp m}|R_n\rangle, \\
|L_n\rangle &= \sum_m \tilde{\alpha}_{nm}|L_m\rangle + \tilde{\beta}_{nm}|L_{n\perp m}\rangle, \qquad \tilde{\alpha}_{nm} = \langle L_m|L_n\rangle, \qquad \tilde{\beta}_{nm} = \langle L_{n\perp m}|L_n\rangle,
\end{align}
with the biorthogonality condition $\langle L_m|R_n\rangle = \delta_{mn}$, and where 
$|R_{n\perp m}\rangle$ represents the component of $|R\rangle_n$ perpendicular to $|R\rangle_m$~\cite{ETH_nh_ours}. 
The result is
\begin{equation}
\mathcal{F}=\sqrt{\text{Tr}(|\hat{H} - \hat{H}^\dagger|^2)} = \sqrt{\sum_{m,n} \left( \varepsilon_m^* \varepsilon_n \alpha_{nm} \tilde{\alpha}_{nm}^* + \varepsilon_n^* \varepsilon_m \tilde{\alpha}_{nm} \alpha_{nm}^* \right) -  \sum_m \left( \varepsilon_m^{*2} + \varepsilon_m^2 \right)}.
\label{eqn:frobenius}
\end{equation}

As this result shows, the Frobenius norm captures both the influence of non-real energy eigenvalues $\varepsilon_n$ as well as non-trivial overlaps $\alpha_{nm}$, $\tilde{\alpha}_{nm}$. 
Figure~\ref{fig:Ham_frobnorm}(a) shows the behavior of $\mathcal{D}^{\prime}$ in the interacting non-Hermitian Hatano--Nelson model. 
Notably, it exhibits qualitatively similar trends to the operator norm $\mathcal{D}$ analyzed in the main text [see Fig.~\ref{fig:Ham}(b)]. 
It is well conceivable that other relevant functions of $\varepsilon_n$, $\alpha_{nm}$, and $\tilde{\alpha}_{nm}$ can be constructed to capture (or even better showcase) the salient features of the system.

As noted earlier,  $\mathcal{D}^{\prime}$ becomes nonzero in approximately the same region as the PT breaking transition. 
In Fig.~\ref{fig:finite-size_scaling}, we present the system-size dependence of $\mathcal{D}^{\prime}$ and compare it with the PT symmetry-breaking transition identified by the nonanalyticity in $\Delta_{01}$, in order to examine in how far both agree in the large-system limit.
In Fig.~\ref{fig:finite-size_scaling}(a), we show the variation of $\mathcal{D}^{\prime}$ and its derivative $\mathrm{d}\mathcal{D}^{\prime}/\mathrm{d}V$, while panel (c) shows $\Delta_{01}$, all as a function of $V$ with fixed $\chi=2.5$. 
The onset of nonanalyticity marks the PT symmetry-breaking transition $\Delta_{01}$. The Hamiltonian non-Hermiticity evolves smoothly across this transition region rather than developing a sharper transition with increasing $N$. 

To further quantify the difference between both observables, we derive the critical point $V^{\mathrm c}_{\Delta}$ (defined as the peak position of $\Delta_{01}$) and the crossover point $V^{\mathrm c}_{\mathcal{D}^{\prime}}$ (given by the minimum of the derivative of ${\mathcal{D}^{\prime}}$), whose system-size scaling is shown in panels (b) and (d).  The two measures converge to distinct values as $N \to \infty$. 
This quantitative difference may perhaps be unsurprising: 
The PT-breaking transition corresponds to an exceptional point between two lowest excited states, whereas the Hamiltonian non-Hermiticity reflects global characteristics of the Hamiltonian across its entire spectrum, as is also evident from Eq.~\eqref{eqn:frobenius}.

Furthermore, to extend our analysis of the non-Hermiticity score of observables to non-local observables, we consider the staggered magnetization  $\hat{I} = \sum_i (-1)^i \hat{n}_i$.
Figure~\ref{fig:Ham_frobnorm}(b) shows the trend of $SC^{\hat{I}}_{\rm max}$ across the $\chi$-$V$
parameter space. This non-local observable exhibits behavior qualitatively similar to the site occupation shown in Fig.~\ref{fig:Hatano_nelson}(a).

\begin{figure}[ht]
  \centering
\includegraphics[width=14cm]{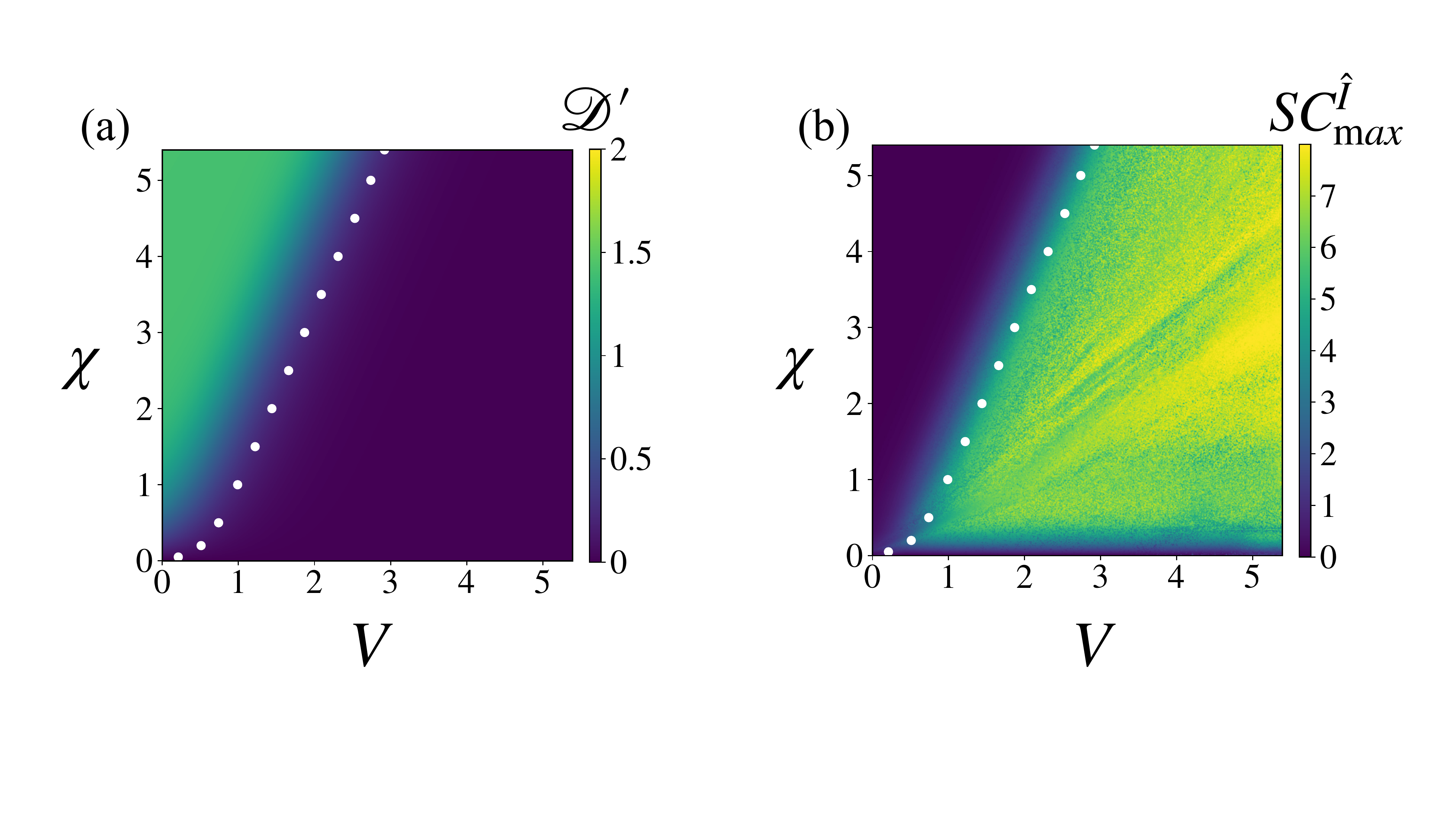}

  \caption{
    (a) Behavior of  $\mathcal{D}'=\frac{||(H^{\mathrm{HN}}_{\mathrm{nh}} - (H^{\mathrm{HN}}_{\mathrm{nh}})^\dagger)||_{\mathrm{F}}}{||H^{\mathrm{HN}}_{\mathrm{nh}}||_{\mathrm{F}}}$, where $||...||_{\mathrm F}$ denotes Frobenius norm,  for the interacting non-Hermitian Hatano--Nelson model. The behavior remains qualitatively same as that obtained in Fig.~\ref{fig:Ham}(b), with all parameters identical to those considered there. (b) Trend of $SC^{\hat{I}}_{\rm max}$, where 
    $\hat{I} = \sum_i (-1)^i \hat{n}_i$,   for the interacting non-Hermitian Hatano--Nelson model  in the $\chi-V$ plane. The behavior remains qualitatively the same as that obtained in Fig.~\ref{fig:Hatano_nelson}(a), which considers identical parameters.
  }
  \label{fig:Ham_frobnorm}
\end{figure}

\begin{figure}[ht]
  \centering
\includegraphics[width=14cm]{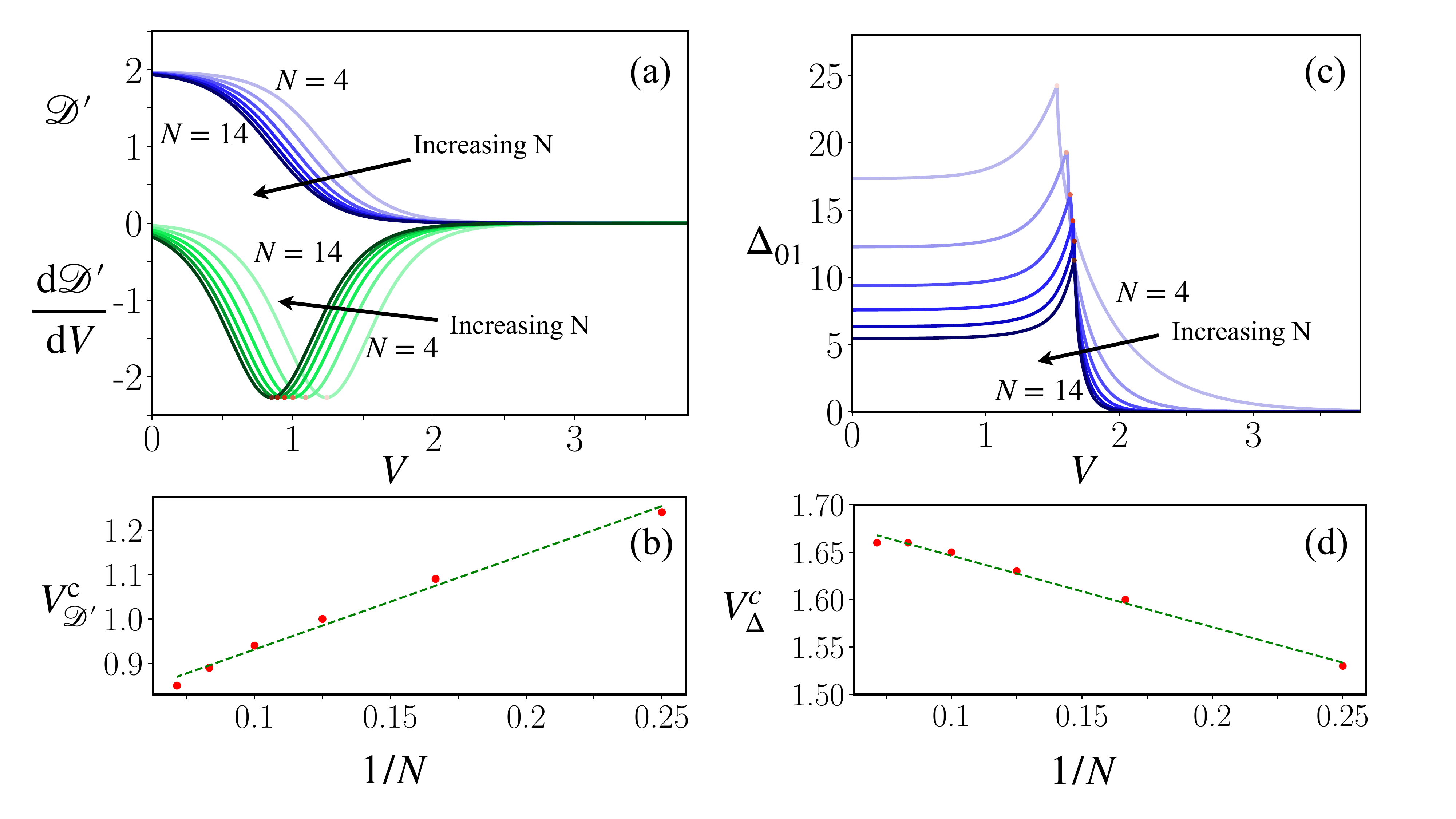}
  \caption{
    (a) Variation of $\mathcal{D}^{\prime}$ and  $\frac{\mathrm{d}\mathcal{D}^{\prime}}{\mathrm{d}V}$ with  respect to $V$ for fixed $\chi=2.5$. The different curves correspond to system sizes $N=4, 6, 8, 10, 12 $ and $14$. We consider the minimum of $\frac{\mathrm{d}\mathcal{D}^{\prime}}{\mathrm{d}V}$ to locate the transition point $V_{\mathcal{D}^{\prime}}^{\rm c}$. (b) Finite-size scaling analysis of $V_{\mathcal{D}^{\prime}}^{\rm c}$. The fit approximates $V_{\mathcal{D}^{\prime}}^{\rm c}\approx 0.72\pm 0.02$ at $N\approx \infty$. (c) Variation of $\Delta_{01}$ with $V$ for different system sizes. The non-analytic point in $\Delta_{01}$ locates the critical point $V^{\rm}_{\Delta}$ for the PT transition. (d) Finite size analysis of $\Delta_{01}^{\rm c}$. The fit approximates $\Delta_{01}^{\rm c}\approx1.72\pm 0.005$ at $N\approx\infty.$ 
    }
  \label{fig:finite-size_scaling}
\end{figure}

\section{Behavior of $\tilde{\mathcal{D}}=|\hat{H}^{\mathrm{HN}}_{\mathrm{nh}}-(\hat{H}^{\mathrm{HN}}_{\mathrm{nh}})^\dagger|$ in the $V-\chi$ plane }
\label{appendixC}
In this appendix, we present numerical results using---rather than the normalized Hamiltonian non-Hermiticity score, Eq.~\eqref{eq:Hnhscore}---the unnormalized norm $\tilde{\mathcal{D}}=|\hat{H}^{\mathrm{HN}}_{\mathrm{nh}}-(\hat{H}^{\mathrm{HN}}_{\mathrm{nh}})^\dagger|$. 
Figure~\ref{fig:Ham_unnorm} shows the results computed for the interacting Hatano--Nelson model. 
As may have been expected, the distance between $\hat{H}^{\mathrm{HN}}_{\mathrm{nh}}$ and its Hermitian conjugate  $(\hat{H}^{\mathrm{HN}}_{\mathrm{nh}})^\dagger$ depends only on the strength of the asymmetric hopping $\chi$. 
As a consequence, the use of the unnormalized form would suggest physics that remains completely independent of the interaction strength $V$. 
In particular, it does not capture in any way the changes that occur in the observable scores as $V$ is increased [see Figs.~\ref{fig:Hatano_nelson} and \ref{fig:score_count} of the main text]. 
Moreover, at large interaction the non-Hermitian part becomes insignificant in comparison to the interactions. 
We thus find it more suitable to normalize by the operator norm $|\hat{H}^{\mathrm{HN}}_{\mathrm{nh}}|$ to obtain a non-Hermitian score $\mathcal{D}$ corresponding to a Hamiltonian with fixed bandwidth [as given in Eq.~\eqref{eq:Hnhscore} and Fig.~\ref{fig:Ham}(b) of the main text].

\begin{figure}[h]
  \centering
  \includegraphics[width=9cm]{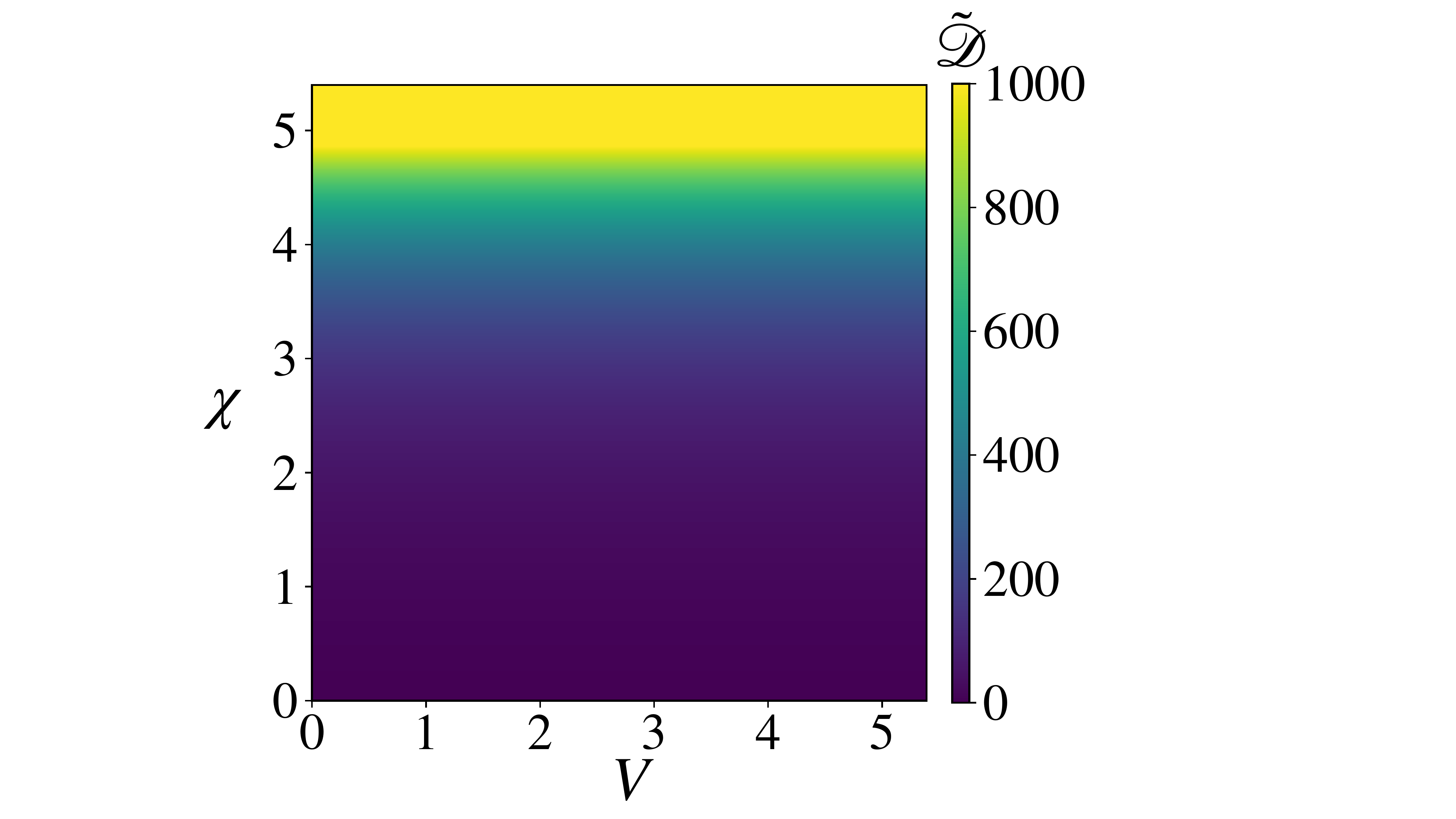}
\caption{Unnormalized degree of non-Hermiticity $\tilde{\mathcal{D}}=||\hat{H}^{\mathrm{HN}}_{\mathrm{nh}}-(\hat{H}^{\mathrm{HN}}_{\mathrm{nh}})^\dagger||$   for  the interacting non-Hermitian Hatano--Nelson model as in Fig.~\ref{fig:Ham}(b). The distance between $\hat{H}^{\mathrm{HN}}_{\mathrm{nh}}$ and its Hermitian conjugate  $(H^{\mathrm{HN}}_{\mathrm{nh}})^\dagger$ remains independent of the interaction term $V$ and simply increases monotonically with the strength of the asymmetric hopping $\chi$. }
\label{fig:Ham_unnorm}    
\end{figure}

\section{Generation of left and right ensembles}
\label{sec:eigenbasis}

As mentioned in the introduction,  to compute physically relevant quantities of any non-Hermitian system, the non-unique choice of basis states often leads to anomalous behavior. In this section, we argue that expectation values evaluated in the right and left eigenvectors of the model have a valid physical origin \cite{ETH_nh_ours,McDonald}. 

Let us consider a non-Hermitian Hamiltonian $H_{\mathrm{nh}}$ expressed in the biorthogonal  basis
\begin{eqnarray}
\hat{H}_{\mathrm{nh}}=\sum_m \lambda_m |R_m\rangle \langle L_m|,
\end{eqnarray}
 where $\langle L_m|R_n \rangle=\delta_{mn}$ and the set of right (left) eigenvectors $\{|R_m\rangle\}$ ($\{\langle L_n|\}$) are not necessarily orthonormal. We now consider the evolution it generates, 
\begin{eqnarray}
\hat{\rho}_{RR}(t)&=&e^{-i\hat{H}_{\mathrm{nh}}t}\hat{\rho}_{\mathrm{in}}e^{it \hat{H}_{\mathrm{nh}}^\dagger}, \nonumber \\
&=&\sum_{i} e^{-i( \lambda_m- \lambda_n^*)t} a_{mn}  |R_m\rangle \langle R_n|,
\label{eqn:dynamics1}
\end{eqnarray}
where $a_{mn}=\langle L_m|\hat{\rho}_{\mathrm{in}}|L_n\rangle$. 
We can distinguish three situations.\\

\noindent 
{\bf 1.   All the  $\lambda_m$'s have non-vanishing imaginary part with a non-degenerate $\max\{\Im[\lambda_{\tilde{m}}]\}=\lambda$.} 

Under this condition,  Eq.~(\ref{eqn:dynamics1}) can be rewritten as
\begin{align}
\hat{\rho}_{RR}(t)&=\sum_m e^{2t \Im[\lambda_m]}a_{mm}|R_m\rangle \langle R_m|\\
  &  +\sum_{m\neq n}e^{-it(\lambda_m- \lambda_n^*)}a_{mn}|R_{m}\rangle \langle R_{n}|.\nonumber
\label{eqn:dynamics2}
\end{align} 
At large times, $\hat{\rho}_{RR}(t)$ eventually converges to the right eigenvector  with slowest decay rate \cite{nh_quantumjump2,McDonald}, 
\begin{eqnarray}
\hat{\rho}_{RR}(t) \xrightarrow{t\rightarrow \infty}  e^{2t \lambda }a_{\tilde{m}\tilde{m}}|R_{\tilde{m}}\rangle \langle R_{\tilde{m}}|.
\end{eqnarray}
In general, physical expectation values are given by $\mathrm{Tr}(\hat{\rho}_{RR}(t)\hat{O})$. Taking the trace in the biorthogonal basis that satisfies the completeness relation $\sum_m |R_m\rangle \langle L_m|=\mathbb{I}$, and using the relation $ \langle R_n|L_k\rangle=\delta_{nk}$, we thus find physical observables will be obtained from the expectation values $\langle R_{\tilde{m}}|\hat{O}|R_{\tilde{m}}\rangle$.\\

\noindent 
{\bf 2. All  $\lambda_m$'s are real (Parity-Time symmetric case).}

 In this case, Eq.~(\ref{eqn:dynamics1}) reduces to  
\begin{align}
\hat{\rho}_{RR}(t) & = \sum_m a_{mm}|R_m\rangle \langle R_m|\nonumber\\  &  + a_{mn}\sum_{m\neq n}e^{-it(\lambda_m-\lambda_n)}|R_{m}\rangle \langle R_{n}|\,.
\end{align}
This situation resembles  the Hermitian case; neglecting resonances, the off-diagonal fluctuations are suppressed at large times, resulting in a diagonal ensemble consisting of the right eigenvectors given by 
\begin{equation}
\hat{\rho}_{RR}(t) \xrightarrow{t\rightarrow \infty} \sum_m a_{mm} |R_{m}\rangle \langle R_{m}|.
\end{equation}
Similar to the previous case, here also the observables at long times are determined by (suitably weighted) matrix elements $\langle R_{m}|\hat{O}|R_{m}\rangle$ in the right eigenbasis.\\

\noindent 
{\bf 3.  Arbitrary non-Hermitian time evolution.}

In the general scenario, an observable $\hat{O}$ undergoes the evolution 
\begin{eqnarray}
\text{Tr}(\hat{O}\hat{\rho}_{RR}(t))
&=&\sum_{mn} e^{-it(\lambda_m-\lambda_n^*)} a_{mn}  \langle R_n|\hat{O}|R_m\rangle\,.
\end{eqnarray}
While the expression is more complicated, the matrix elements of the observable are still determined by the right eigenvectors, while the left eigenvectors enter through the overlap with the initial state via $a_{mn}$.\\

All the scenarios considered above illustrate the physical origin of expectation values evaluated in the right eigenvectors of the model. A similar analysis with $\hat{H}_{\mathrm{nh}}^\dagger$ justifies the same reasoning for the left eigenvectors ($\{\langle L_m|\}$). As these considerations also show, while possibly containing interesting properties of the model \cite{McDonald}, terms like $\langle R_m|\hat{O}|L_n\rangle$ do not naturally appear in the evaluation of physical quantities for the quantum state evolved under any generic non-Hermitian dynamics.

\section{Derivation of analytical of magnetization and entanglement for imperfect Bell states}
\label{appendixA}
In this appendix, we present the analytical form of the magnetization and quantum entanglement computed for the set of left and right eigenvectors of the non-Hermitian Hamiltonian  $\hat{H}_{\mathrm{nh}}^\mathrm{Bell}$. Values of $m_z$ evaluated for the ensembles of right and left eigenvectors of $\hat{H}_{\mathrm{nh}}^\mathrm{Bell}$ are given in the following table:  
\begin{center}
\begin{tabular}{||c |c| c ||} 
 \hline
 Index ($m$)& $m_z=\text{Tr}(\hat{\sigma}_z \hat{\rho}_{m}^L)$,   $\hat{\rho}_{m}^L=\text{Tr}_1[|L_m\rangle \langle L_m|]$ & $m_z=\text{Tr}(\hat{\sigma}_z \hat{\rho}_{m}^R)$, $\hat{\rho}_{m}^R=\text{Tr}_1[|R_m\rangle \langle R_m|]$ \\ [0.5ex] 
 \hline\hline
 1 & $\frac{\alpha^2+\alpha^2(\alpha-1)^2-1}{\alpha^2+\alpha^2(\alpha-1)^2+1}$ &  $\frac{1-\alpha^2}{1+\alpha^2}$ \\ 
 \hline
 2 & $\frac{\alpha^2+\alpha^2(\alpha-1)^2-1}{\alpha^2+\alpha^2(\alpha-1)^2+1}$  & $\frac{1-\alpha^2}{1+\alpha^2}$  \\
 \hline
 3 & 0 &  $\frac{(1-\alpha)^2}{2+(1-\alpha)^2}$  \\
 \hline
 4 & 0 & $\frac{(1-\alpha)^2}{2+(1-\alpha)^2}$\\ [1ex] 
 \hline
\end{tabular}
\end{center}

We can write the entanglement entropy as $\mathcal{S}(|R_m\rangle\langle R_m|)=-\lambda_m^R \log(\lambda_m^R)-(1-\lambda_m^R) \log(1-\lambda_m^R)$ and $\mathcal{S}(|L_m\rangle\langle L_m|)=-\lambda_m^L \log(\lambda_m^L)-(1-\lambda_m^L) \log(1-\lambda_m^L)$. Analytical forms of the quantities $\lambda_m^R$ and $\lambda_m^L$ are:  
\begin{center}
\begin{tabular}{||c |c| c ||} 
 \hline
 Index (m)&    $\lambda_m^L$ & $\lambda_m^R$ \\ [0.5ex] 
 \hline\hline
 1 & $\frac{1}{2}+\frac{\sqrt{1-\frac{4\alpha^2}{(\alpha^2+\alpha^2(\alpha-1)^2+1)^2}}}{2}$  &  $\frac{1}{1+\alpha^2}$ \\ 
 \hline
 2 & $\frac{1}{2}+\frac{\sqrt{1-\frac{4\alpha^2}{(\alpha^2+\alpha^2(\alpha-1)^2+1)^2}}}{2}$  & $\frac{1}{1+\alpha^2}$  \\
 \hline
 3 &   $\frac{1}{2}$  &  $\frac{1}{2}+\frac{\sqrt{1-\frac{4}{(2+(\alpha-1)^2)^2}}}{2}$ \\
 \hline
 4 & $\frac{1}{2}$   & $\frac{1}{2}+\frac{\sqrt{1-\frac{4}{(2+(\alpha-1)^2)^2}}}{2}$\\ [1ex] 
 \hline
\end{tabular}
\end{center}

\section{Purity score for maximally mixed input state}
\label{appendixB}
In this section, we give the analytical derivation for the purity score of the model given by imperfect Bell states. The purity of the ensemble evolved under $\hat{H}_{\mathrm{nh}}^\mathrm{Bell}$ is 
\begin{equation}
\mathcal{P}(\hat{\rho}_{RR}(t))=\frac{\text{Tr}\Big[\hat{\rho}_{RR}(t)^2\Big]}{\text{Tr}(\hat{\rho}_{RR}(t))^2}=\frac{\sum_{mn  m'n'}p_{mn}p_{m'n'}\langle R_{n'}|R_m\rangle \langle R_n|R_{m'}\rangle}{\Big(\sum_{mn}p_{mn}\langle R_{n}|R_m\rangle\Big)^2}\,,
\label{purity_right}
\end{equation}
with $p_{mn}=e^{-i( \lambda_m- \lambda_n)t}  \langle L_m|\hat{\rho}_{\mathrm{in}}|L_n\rangle$.
Similarly, for the quantum state evolved under $(\hat{H}^{\mathrm{Bell}}_{\mathrm{nh}})^\dagger$, we get
\begin{equation}
\mathcal{P}(\hat{\rho}_{LL}(t))=\frac{\text{Tr}\Big[\hat{\rho}_{LL}(t)^2\Big]}{\text{Tr}(\hat{\rho}_{LL}(t))^2}=\frac{\sum_{mn,m'n'}{\tilde p}_{mn} {\tilde p}_{m'n'}\langle L_{n'}|L_m\rangle \langle L_n|L_{m'}\rangle|}{\Big(\sum_{mn}\tilde{p}_{mn}\langle L_{n}|L_m\rangle\Big)^2}\,,
\label{purity_left}
\end{equation}
with $\tilde{p}_{mn}=e^{-i( \lambda_m- \lambda_n)t}  \langle R_m|\hat{\rho}_{\mathrm{in}}|R_n\rangle$. 

Now, for a maximally mixed initial state $\hat{\rho}=\frac{\mathbb{I}}{4}$, from Eq.~(\ref{purity_right}) we get
\begin{eqnarray}
\mathcal{P}(\hat{\rho}_{RR}(t))&=&\frac{\sum_{mn  m'n'} \langle L_m|L_n\rangle  \langle L_{m'}|L_{n'}\rangle  \langle R_{n'}|R_m\rangle \langle R_n|R_{m'}\rangle}{\Big(\sum_{mn} \langle L_{m}|L_n\rangle\langle R_{n}|R_m\rangle\Big)^2}\nonumber\\
&=& \frac{\sum_{mn  m'n'} \langle L_n|L_m\rangle  \langle L_{n'}|L_{m'}\rangle  \langle R_{m}|R_{n'}\rangle \langle R_{m'}|R_{n}\rangle}{\Big(\sum_{mn} \langle L_{n}|L_m\rangle\langle R_{m}|R_n\rangle\Big)^2} \ \text{(taking the conjugate of all the overlaps)}\nonumber\\
&=& \frac{\sum_{mn  m'n'} \langle R_{m}|R_{n}\rangle \langle R_{m'}|R_{n'}\rangle \langle L_{n'}|L_m\rangle  \langle L_{n}|L_{m'}\rangle  }{\Big(\sum_{mn} \langle L_{n}|L_m\rangle\langle R_{m}|R_n\rangle\Big)^2} \ \text{(exchanging indices $n$ and $n'$)}\nonumber\\
&=&\mathcal{P}(\hat{\rho}_{LL}(t)).
\end{eqnarray}
Hence, for a maximally mixed initial state, the purity obtained at any time remains exactly the same for the left and right ensembles and thus the non-Hermitcity score for the purity vanishes.

\section{Operational  interpretation of dynamics under any non-Hermitian  evolution}
\label{sec:kinematic} 
We start our discussion by providing an operational interpretation of dynamics governed by any generic non-Hermitian  Hamiltonian $\hat{H}_{\mathrm{nh}}$. 
Let us consider a general quantum state 
\begin{eqnarray}
\hat{\rho}(t)=\sum_\lambda p_\lambda \hat{P}_\lambda,
\label{eq:state}
\end{eqnarray} 
where $\hat{P}_\lambda$ is  a non-negative operator, $\hat{P}_\lambda=\hat{L}_\lambda^\dagger \hat{L}_\lambda$, and $p_\lambda \geq 0$ and $\sum_\lambda p_\lambda=1$. 
Now, a dynamical evolution of the above state under any  map \cite{book1} would result in
\begin{equation}
\label{eq:krausmap}
\hat{\rho}(t+\delta t)=\sum_{m} \hat{\mathcal{K}}_m(\delta t) \hat{\rho}(t)\hat{\mathcal{K}}^\dagger_m(\delta t)=\sum_\lambda p_\lambda \hat{\tilde{P}}_\lambda,
\end{equation}
where $\sum_m \hat{\mathcal{K}}^\dagger_m \hat{\mathcal{K}}_m=\mathbb{I}$ and  $\hat{\tilde{P}}_\lambda=\sum_{m}\hat{\mathcal{K}}_m(\delta t) \hat{L}^\dagger_\lambda \hat{L}_\lambda \hat{\mathcal{K}}^\dagger_m(\delta t)$ is again a positive operator (because each summand $\hat{\mathcal{K}}_m \hat{L}^\dagger_\lambda \hat{L}_\lambda \hat{\mathcal{K}}^\dagger_m$ is), with $\hat{\tilde{P}}_\lambda \geq 0$ and $\sum_\lambda \hat{\tilde{P}}_\lambda=\mathbb{I}$. 

We now consider a constrained evolution by projecting the evolved state $\hat{\rho}(t+\delta t)$ onto a subspace,
 \begin{equation}
\hat{\tilde{\rho}}(t+\delta t)=\hat{\mathcal{P}}\hat{\rho}(t+\delta t) \hat{\mathcal{P}}=\sum_\lambda p_\lambda   \hat{\mathcal{M}}_\lambda\,,
\label{eq:evol}
\end{equation}
where $\hat{\mathcal{M}}_\lambda=\hat{\mathcal{P}} \hat{\tilde{P}}_\lambda \hat{\mathcal{P}}$. 
In the case where $\hat{\mathcal{P}}\neq \mathbb{I}$ (i.e., a projector onto a true subspace of the full Hilbert space), we have $\sum_\lambda   \hat{\mathcal{M}}_\lambda \neq \mathbb{I}$. 
The subspace restriction thus can lead to a loss of norm,  $\Tr(\hat{\tilde{\rho}}(t+\delta t)) \leq 1$. 
Nevertheless, $\hat{\tilde{\rho}}(t+\delta t)$ satisfies all the necessary conditions for a valid quantum state in the  language of kinematic evolution:
\begin{itemize}
\item[] (i) {\bf Positive operator:} We have $\hat{\mathcal{M}}_\lambda \geq 0$ and $p_\lambda\ge0$. This implies $\hat{\tilde{\rho}}(t+\delta t) \geq0$. 
\item[] (ii) {\bf Positive measurement probabilities:} If we measure some quantum observable $\hat{O}=\sum_\ell a_\ell \hat{O}_\ell$, the probabilities of distinct outcomes $a_\ell$ are positive, since using (i) we have $p_\ell\equiv\Tr(\hat{O}_\ell \hat{\tilde{\rho}}(t+\delta t) \hat{O}_\ell )\geq 0$. 
\end{itemize}
Thus, while $\sum_\ell p_\ell = \Tr(\hat{\tilde{\rho}}(t+\delta t)) \leq 1$ is possible, probabilities of obtaining measurement results are always positive as expected from a physical observable. The remaining probability is associated to the cases where the system has leaked out into the subspace not considered (given by projector $\mathbb{I}-\hat{\mathcal{P}}$). 
Hence, one can always assign an ``operational meaning" or dynamical interpretation to any quantum state that satisfies the above two conditions.  
Before computing a quantity related to the system, it can be useful to normalize the quantum state $\hat{\tilde{\rho}}(t+\delta t)$, as we do throughout this paper, so as to obtain the probabilities conditioned on the system being in the subspace, which sum up to identity.

We now show that the evolution under an arbitrary non-Hermitian Hamiltonian $\hat{H}_\mathrm{nh}$, defined on the system's Hilbert space $\mathcal{H}_\mathrm{S}$, can be written as a Krauss map acting on an extended space  including an ancilla, $\mathcal{H}_\mathrm{S}\otimes \mathcal{H}_\mathrm{A}$, followed by a suitable projection. I.e., the evolution under any non-Hermitian Hamiltonian takes the form given by Eq.~\eqref{eq:evol}.  
The ancilla can be considered as any generic system ranging from qubits to bosonic baths. For the illustration purpose, here we consider it to be a qubit.

We start by constructing the evolution to leading order in a small time step $\delta t$, from which we obtain the continuous time evolution in the limit $\delta t\to 0$. 
Given any $\hat{H}_\mathrm{nh}$, one can always define the operators $\hat{\mathcal{K}}_{0}=(\mathbb{I}_\mathrm{S}-i\hat{H}_\mathrm{nh}\delta t)\otimes\mathbb{I}_\mathrm{A}$ and $\hat{\mathcal{K}}_{1}=\sqrt{\delta t(\hat{H}_\mathrm{nh}^{\dagger}-\hat{H}_\mathrm{nh}})\otimes\hat{\sigma}_\mathrm{A}^x$ acting in $\mathcal{H}_{\mathrm{S}}\otimes\mathcal{H}_{\mathrm{A}}$. Here, $\hat{\sigma}_\mathrm{A}^x$ is the Pauli matrix and $\hat{\mathcal{K}}_{0}^\dagger\hat{\mathcal{K}}_{0}+\hat{\mathcal{K}}_{1}^\dagger\hat{\mathcal{K}}_{1}= \mathbb{I} + \mathcal{O}(\delta t^2)$. The dynamical evolution of an initial state operator $\hat{\rho}(t)$ acting on $\mathcal{H}_{\mathrm{S}}\otimes\mathcal{H}_{\mathrm{A}}$ is given by 
\begin{equation}
\hat{\rho}(t+\delta t) = \hat{\mathcal{K}}_0 \hat{\rho}(t) \hat{\mathcal{K}}_0^\dagger + \hat{\mathcal{K}}_1 \hat{\rho}(t)\hat{\mathcal{K}}_1^\dagger\,.
\end{equation}
This evolution describes a Master equation with jump operators $\hat{\mathcal{K}}_1$ and is in the Krauss form as given in Eq.~\eqref{eq:krausmap}. (Note that the considered form of the operators  $\hat{\mathcal{K}}_0$ and $\hat{\mathcal{K}}_1$ is not unique; in particular, a specific physical implementation yielding the same $\hat{H}_\mathrm{nh}$ may be described microscopically by a different set of Krauss or unitary operators
\cite{dilation1, dilation2, dilation3, dilation4, dilation5, dilation6}.) 

To obtain the desired evolution, we assume the ancilla qubit to be initialized in the $|0\rangle\langle 0|$ state. 
We then employ dynamical evolution under the above map. Afterwards, we project the evolved state back onto $\hat{\mathcal{P}}_0 = |0\rangle\langle 0|$. 
This amounts to using the qubit as a ``flag'' permitting postselection onto the no-jump trajectory \cite{nh_quantumjump2,daley_quat_trajec}. 
The result is
\begin{eqnarray}
\hat{\mathcal{P}}_0 \hat{\rho}(t+\delta t) \hat{\mathcal{P}}_0 
&=& \hat{\mathcal{P}}_0 \hat{\mathcal{K}}_0 (\hat{\rho}_\mathrm{S}(t)\otimes|0\rangle\langle0|) \hat{\mathcal{K}}_0^\dagger \hat{\mathcal{P}}_0 \nonumber\\&&+ \hat{\mathcal{P}}_0 \hat{\mathcal{K}}_1 (\hat{\rho}_\mathrm{S}(t)\otimes|0\rangle\langle0|)\hat{\mathcal{K}}_1^\dagger\hat{\mathcal{P}}_0\nonumber\\
&=& (\mathbb{I}_\mathrm{S}-i\hat{H}_\mathrm{nh}\delta t)\hat{\rho}_\mathrm{S}(t)(\mathbb{I}_\mathrm{S}+i\hat{H}_\mathrm{nh}^{\dagger}\delta t)\otimes|0\rangle\langle0|\nonumber\\
&&+\mathcal{O}(\delta t^2).
\end{eqnarray}
Therefore, in the limit of a small time step $\delta t$, the system undergoes the non-Hermitian evolution given by $e^{-it\hat{H}_{\mathrm{nh}}}\hat{\rho}_{\mathrm{S}} e^{it\hat{H}^\dagger_{\mathrm{nh}}}$, whereas the ancilla remains in the initialized $|0\rangle\langle 0|$ state. 
One can see this evolution as a quantum-Zeno effect, generated by the continuous monitoring described by $\hat{\mathcal{P}}_0$, and which freezes the ancilla to the $|0\rangle\langle 0|$ state. As is typical for the quantum-Zeno effect, the system undergoes a non-Hermitian evolution, describing the loss of norm out of the Zeno subspace \cite{kevinphilipp,qzeno1,qzeno2}.

In summary, if we consider a quantum system evolved under the effect of an {\it arbitrary}  non-Hermitian system, we can always assign an operational meaning to the evolved states: they can be understood as {\it quantum states originated due to the action of a constrained dynamical evolution manifested using complete quantum measurement and postselection of the results}. 
This interpretation shows that the resulting time-evolved state has a valid probability distribution, meaning observables can be measured and a physical meaning can be associated to them. Note that from a known implementation of a non-Hermitian Hamiltonian $H_{\mathrm{nh}}$ from an open-system dynamics, it may not always be straightforward to realize also $\hat{H}^\dagger_{\mathrm{nh}}$. In the following section, we give an explicit example for the case of the Hatano--Nelson model that shows how to  effectively realize both $\hat{H}_{\mathrm{nh}}$ and $\hat{H}^\dagger_{\mathrm{nh}}$.

\section{Realization of non-Hermitian Hatano--Nelson model ($\hat{H}_{\mathrm{nh}}^{\mathrm{HN}}$  and  $(\hat{H}_{\mathrm{nh}}^{\mathrm{HN}})^\dagger$) through open quantum system dynamics}
\label{sec:hatanorealization}
In this section, we show how to engineer the jump operators, following approaches outlined in Refs.~\cite{Gong_2018, kevinphilipp}, to  effectively realize both \(\hat{H}_{\mathrm{nh}}^{\mathrm{HN}}\) and its adjoint \((\hat{H}_{\mathrm{nh}}^{\mathrm{HN}})^\dagger\) of the non-Hermitian Hatano--Nelson model (after dropping an overall background loss that does not influence the dynamics). 
Consider  a dissipative dynamics by coupling the system to a large Markovian environment, which yields the following Lindblad master equation
\begin{equation}
\partial_{t}\hat{\rho} = -i[\hat{H}, \hat{\rho}] + \sum_{m}\gamma_{m}\left[\hat{\mathcal{K}}_{m}\hat{\rho}\hat{\mathcal{K}}_{m}^{\dagger} - \frac{1}{2}\{\hat{\mathcal{K}}^{\dagger}_{m}\hat{\mathcal{K}}_{m}, \hat{\rho}\} \right],
\label{eq:open_quantum}
\end{equation}
where $\hat{\rho}$ is the density matrix of the system, $ \hat{\mathcal{K}}_{m} $ are the jump operators for $ m $ different channels, $ \gamma_{m} $ are the jump strengths, and $ [\bullet] $ ($ \{\bullet\} $) denote the commutation (anti-commutation) operation. Here, the first term on the right-hand side corresponds to the Hermitian evolution of the system under the Hamiltonian $ H $. The equation can be reformulated as
\begin{equation}
\partial_{t}\hat{\rho} = -i[\hat{H}_{\rm nh}\hat{\rho}-\hat{\rho} \hat{H}_{\rm nh}^{\dagger}] + \sum_{m}\gamma_{m}\hat{\mathcal{K}}_{m}\hat{\rho}\hat{\mathcal{K}}_{m}^{\dagger},
\end{equation}
where $\hat{H}_{\rm nh} = \hat{H} - \frac{i}{2}\sum_{m}\gamma_{m}\hat{\mathcal{K}}_{m}^{\dagger}\hat{\mathcal{K}}_{m}$ corresponds to a non-Hermitian Hamiltonian. 

To arrive at the specific case considered in our work -- the interacting Hatano-Nelson model Hamiltonian, $\hat{H}_{\rm nh}^{\rm HN}$, as given in Eq.~\eqref{eq:hatano_nelson} -- 
we consider a dissipative system governed by the Hermitian operator
\begin{equation}
 \hat{H}=\sum_i^N\big[ -J\cosh{(\chi)}( \hat{c}^\dagger_i\hat{c}_{i+1}+ \hat{c}^\dagger_{i+1} \hat{c}_{i})+V \hat{n}_i \hat{n}_{i+1} \big],\,
\end{equation} 
and subject to non-local one-body loss processes described by the jump operators
\begin{equation}
\label{eq:jump1}
\hat{\mathcal{K}}_i = \hat{c}_i - i\hat{c}_{i+1},
\end{equation}
with uniform dissipation strength $\gamma_{i}=2J\sinh{(\chi)}$ for all $i$. Then, in the no-jump limit, the system evolves under the non-Hermitian Hamiltonian
\begin{align}
\hat{H}_{\rm nh} &= \sum_{i}\Big[-J\cosh{(\chi)}( \hat{c}^\dagger_i\hat{c}_{i+1}+ \hat{c}^\dagger_{i+1} \hat{c}_{i})+V \hat{n}_i \hat{n}_{i+1}\Big] -\frac{i}{2}\sum_{i}2J\sinh{(\chi)}(\hat{c}^{\dagger}_i + i\hat{c}^{\dagger}_{i+1})(\hat{c}_i - i\hat{c}_{i+1})\nonumber\\
&=\hat{H}_{\rm nh}^{\rm HN}-i2J\sinh{(\chi)\hat{N}},
\end{align}
where $\hat{N}=\sum_{i}\hat{n}_{i}$ denotes the total particle number. 
Now, we replace the non-local one-body loss operators by
\begin{equation}
\label{eq:jump2}
\hat{\tilde{\mathcal{K}}}_i = \hat{c}_i +i\hat{c}_{i+1},
\end{equation}
which in the no-click limit leads to a non-Hermitian evolution under
\begin{align}
\hat{H}_{\rm nh}' &= \sum_{i}\Big[-J\cosh{(\chi)}( \hat{c}^\dagger_i\hat{c}_{i+1}+ \hat{c}^\dagger_{i+1} \hat{c}_{i})+V \hat{n}_i \hat{n}_{i+1}\Big] -\frac{i}{2}\sum_{i}2J\sinh{(\chi)}(\hat{c}^{\dagger}_i - i\hat{c}^{\dagger}_{i+1})(\hat{c}_i + i\hat{c}_{i+1})\nonumber\\
&=(\hat{H}_{\rm nh}^{\rm HN})^{\dagger}-i2J\sinh{(\chi)\hat{N}}.
\end{align}

Although $\hat{H}_{\rm nh}'\neq (\hat{H}_{\rm nh})^\dagger$, the difference arises only from the final term, corresponding to a uniform imaginary energy shift, which can be dropped as it does not affect the system's dynamics. This can be seen as follows: 
Since $[\hat{N}, \hat{H}_{\rm nh}]= [\hat{N}, \hat{H}_{\rm nh}^{\prime}] =0$, the time evolution of any initial wave function can be written as
\begin{equation}
e^{-i\hat{H}_{\rm nh}t} |\psi\rangle = e^{-i\hat{H}^{\rm HN}_{\rm nh}t} e^{-2J\sinh{(\chi)}\hat{N}t}|\psi\rangle =  e^{-i\hat{H}^{\rm HN}_{\rm nh}t} |\tilde{\psi}\rangle
\end{equation}
and 
\begin{equation}
e^{-i\hat{H}_{\rm nh}^{\prime}t} |\psi\rangle = e^{-i(\hat{H}^{\rm HN}_{\rm nh})^{\dagger}t} e^{-2J\sinh{(\chi)}\hat{N}t}|\psi\rangle =  e^{-i(\hat{H}^{\rm HN}_{\rm nh})^{\dagger}t} |\tilde{\psi}\rangle.
\end{equation}
For an initial state with a definite particle number $N$, we have $\exp[-2J\sinh(\chi)\hat{N} t] = \exp[-2J\sinh(\chi) N t],$
which can be dropped (under appropriate renormalization of observables) since it only contributes an overall background loss that does not influence the system's dynamics.

In this way, an effective non-Hermitian evolution under $\hat{H}_{\rm nh}^{\rm HN}$ and $(\hat{H}_{\rm nh}^{\rm HN})^{\dagger}$ is obtained through engineering the loss processes.
Reference~\cite{Gong_2018} presents a protocol for realizing the Hatano--Nelson model in a dissipative optical lattice using cold atoms. In this setup, non-local loss processes---described by the jump operators in Eqs.~\eqref{eq:jump1} and~\eqref{eq:jump2}---are engineered via Rabi couplings induced by an additional running-wave laser. The required relative phase between the operators $\hat{c}_i$ and $\hat{c}_{i+1}$, as dictated by the structure of the jump operators, can be controlled by adjusting the propagation direction of the running wave relative to the optical lattice.
Alternatively, the jump operators can also be realized following the procedure detailed in Ref.~\cite{kevinphilipp}. In this scheme, sites $i$ and $i+1$ are coherently coupled to the same ancillary state, which in turn is subject to decoherence. Tracing out the ancillary state realizes an effective open-system dynamics on the remaining system. The phases between $\hat{c}_i$ and $\hat{c}_{i+1}$ can be adjusted through the relative phases of the coherent out-couplings.

\end{widetext}

\bibliography{reference}

\end{document}